\documentclass[12pt]{article}

\usepackage{booktabs}
\usepackage{tabulary}
\usepackage{multirow}
\usepackage{rotating}
\usepackage{epsfig}
\usepackage{psfrag}
\usepackage{latexsym}
\usepackage{indentfirst}
\usepackage{fancyhdr}
\usepackage{dsfont}
\usepackage{adjustbox}
\usepackage{amsmath}
\usepackage{amssymb}
\usepackage{amsfonts}
\usepackage{mathrsfs}
\usepackage{amsthm}
\usepackage{pifont}
\usepackage{dsfont}
\usepackage{multirow}
\usepackage{array}
\usepackage{chngpage}
\usepackage{longtable}
\usepackage{cite}
\usepackage{bbold}
\usepackage{color}
\usepackage{braket}
\usepackage{colordvi}
\usepackage{fancybox}
\usepackage[footnotesize]{caption2}
\usepackage{graphicx}
\usepackage[center,footnotesize,hang]{subfigure}
\usepackage{bm}
\usepackage{bbm}
\usepackage{url}
\usepackage[table]{xcolor}
\usepackage[colorlinks, linkcolor=red, anchorcolor=black, citecolor=green]{hyperref}
\usepackage{multirow}
\usepackage{harpoon}
\usepackage{textcomp}  %  for the command \textlbrackdbl
\usepackage{enumitem}
\usepackage{mathrsfs}  % $\mathscr{ABCDEFGHIJKLMNOPQRSTUVWXYZ}$
\usepackage[normalem]{ulem}

\usepackage{tikz}
\usepackage{diagbox}
\usepackage{enumitem}

\newcommand{\PreserveBackslash}[1]{\let\E^c=\\#1\let\\=\E^c}
\newcolumntype{C}[1]{>{\PreserveBackslash\centering}p{#1}}
\newcolumntype{R}[1]{>{\PreserveBackslash\raggedleft}p{#1}}
\newcolumntype{L}[1]{>{\PreserveBackslash\raggedright}p{#1}}
\allowdisplaybreaks  % automated pagination

\newcommand{\bq}{\begin{eqnarray}}
\newcommand{\nq}{\end{eqnarray}}

\newcommand{\ignore}[1]{}

\numberwithin{equation}{section}
\begin{document}
\title{ {\Large \bf
Pati-Salam models with $A_4$ modular symmetry
\\[2mm]}
%{\Large \bf Combining Pati-Salam and modular symmetry\\[2mm]}
}

\date{}

\author{
Gui-Jun~Ding$^{a}$\footnote{E-mail: {\tt
dinggj@ustc.edu.cn}},  \
Si-Yi~Jiang$^{a}$\footnote{E-mail: {\tt
siichiang@mail.ustc.edu.cn}},  \
Stephen~F.~King$^{b}$\footnote{E-mail: {\tt king@soton.ac.uk}}, \
Jun-Nan Lu$^{a}$\footnote{E-mail: {\tt
hitman@mail.ustc.edu.cn}},  \
Bu-Yao Qu$^{a}$\footnote{E-mail: {\tt
qubuyao@mail.ustc.edu.cn}},  \
\\*[20pt]
\centerline{
\begin{minipage}{\linewidth}
\begin{center}
$^a${\it \small
Department of Modern Physics, University of Science and Technology of China,\\
Hefei, Anhui 230026, China}\\[2mm]
$^b${\it \small
Physics and Astronomy, University of Southampton, Southampton, SO17 1BJ, U.K.}
\end{center}
\end{minipage}}
\\[10mm]}
\maketitle
\thispagestyle{empty}

\begin{abstract}

The flavor structure of quarks and leptons and quark-lepton unification are studied in the framework of Pati-Salam models with $A_4$ modular symmetry. The three generations of the left-handed and right-handed fermions are assigned to be triplet or singlets of $A_4$. The light neutrino masses are generated through the type-I seesaw mechanism. We perform a systematic classification of Pati-Salam models according to the transformations of matter fields under the $A_4$ modular symmetry, and the general form of the fermion mass matrix is given. We present four phenomenologically viable benchmark models which provide excellent descriptions of masses and flavor mixing of quarks and leptons, including neutrinos. In such models we find that the normal ordered neutrino mass spectrum is preferred over the inverted case, with neutrinoless double beta decay predicted to be too small to be observed by the next generation of experiments.

\end{abstract}
\newpage

\section{\label{sec:introduction}Introduction}

Understanding the observed pattern of the quark and lepton masses, mixing angles and CP violating phases is one of the most fascinating challenges in particle physics. In the standard model (SM), the hierarchical quark masses, small quark mixing angles and the quark CP violation phases arise from the Yukawa couplings of the quark fields to the SM Higgs field~\cite{Workman:2022ynf}. Analogously the charged lepton Yukawa couplings lead to a mass hierarchy similar to that of the down-type quarks. However, the SM offers no insight into the nature of the Yukawa couplings which are not subject to any constraint from the SM gauge symmetry. The discovery of neutrino oscillation shows that neutrinos have mass and the lepton mixing is drastically different from the quark mixing. Global analysis of neutrino oscillation data shows that the atmospheric mixing angle $\theta^{\ell}_{23}$ and the solar mixing angle $\theta^{\ell}_{12}$ are large, while the reactor angle $\theta^{\ell}_{13}$ is of a similar size to the Cabibbo angle~\cite{Esteban:2020cvm}. Hence the developement of neutrino physics enriches the flavor puzzle, raises new questions such as the disparity of the quark and lepton mixing and the origin of tiny neutrino mass. The masses, mixing angles and CP violating phases of quarks and leptons are all determined from experiments, while the organizing principle behind them is a longstanding open question in particle physics.

Symmetry such as flavor symmetry and the gauge symmetry of the Grand Unified Theory (GUT) turns out to be very useful tool to address the above flavor puzzle of the SM. The flavor symmetry acts on the gauge multiplets of the same species but different generations, consequently it leads to strong correlation between the matrix elements of a given fermion mass matrix~\cite{King:2013eh,Xing:2020ijf,Feruglio:2019ybq,Ding:2024ozt}. On the other hand, the gauge symmetry of GUT relates the different fields within the GUT multiplets, thus connecting the mass matrices of leptons and quarks in a non-trivial way~\cite{King:2017guk}. Hence both flavor symmetry and GUT allow to reduce the number of flavor parameters and consequently lead to predictions which could be tested by experiments. It is natural to combine GUT with flavor symmetry to construct a theory of flavor involving a few free parameters. Since GUT unifies the quarks and leptons within each generation into very few multiplets, the masses and mixing patterns of both quark and lepton should be reproduced in a single model merging GUT and flavor symmetry~\cite{King:2017guk}.

Flavor symmetry can not be exact and it has to be broken by suitably chosen vacuum expectation values (VEVs) of scalar flavon fields, the dynamics achieving the vacuum alignment of flavons considerably complicating the flavor models, since the additional auxiliary symmetry and fields are generally necessary to cleverly design the flavon energy density and get the desired vacuum configuration. Modular invariance as the candidate flavor symmetry was suggested to overcome the drawback of complicated vacuum alignment in traditional flavor symmetry~\cite{Feruglio:2017spp}. Since there is no flavon besides the complex modulus $\tau$ in the simplest implementation, the VEV of $\tau$ being the unique source of flavor symmetry breaking, the vacuum alignment problem is simplified significantly. In the past years, the modular symmetry has been intensively studied from both top-down and bottom-up approaches, see Refs.~\cite{Kobayashi:2023zzc,Ding:2023htn} for recent review. As expected, imposition of modular symmetry on GUTs gives rise to very economical and predictive flavor models, for example the $SU(5)$ GUT~\cite{deAnda:2018ecu,Kobayashi:2019rzp,Du:2020ylx,Zhao:2021jxg,Chen:2021zty,King:2021fhl,Ding:2021zbg,Abe:2023dvr,deMedeirosVarzielas:2023ujt}, flipped $SU(5)\times U(1)$ GUT~\cite{Charalampous:2021gmf,Du:2022lij} and $SO(10)$ GUT~\cite{Ding:2021eva,Ding:2022bzs} in combination with the modular symmetry $\Gamma_2\cong S_3$, $\Gamma_3\cong A_4$, $\Gamma_4\cong S_4$ and $\Gamma'_6\cong S_3\times T'$ have been studied.

In this paper we shall study the earliest proposal for quark-lepton unification, namely the Pati-Salam (PS) model based on the gauge symmetry $SU(4)_C\times SU(2)_L\times SU(2)_R$~
\cite{Pati:1974yy}, in the framework of $A_4$ modular symmetry. This minimal PS unification framework, which includes neutrino mass, combined with the minimal $A_4$ modular group, which includes triplet representations, has not so far been studied in the literature, although many PS models with based on the traditional discrete flavor symmetry have been constructed, for example in Refs.~\cite{deAdelhartToorop:2010vtu,King:2013hoa,King:2014iia,CarcamoHernandez:2017owh}. Moreover, PS is better motivated than $SO(10)$ from the point of view of string theory, since it may be broken by simple Higgs representations which readily arise in string constructions~\cite{Antoniadis:1990hb,Kobayashi:2004ya,Li:2022cqk} rather than the large representations often used in $SO(10)$ which are difficult to obtain from string theory~\cite{Dienes:1996wx}. Modular symmetry is also motivated by heterotic string theory~\cite{Baur:2019kwi}. Indeed, since both modular symmetry and PS are motivated from top-down string constructions, this increases the motivation to study them together in bottom-up constructions. In comparison with the PS models based on traditional flavor symmetries, in the modular symmetry approach followed here the flavon fields will be replaced by modular forms  which renders the theory much more simpler.

In Pati-Salam theories, the SM matter fields such as the quark doublets, two quark singlets, the lepton doublet and the charged lepton singlet plus the right-handed neutrino in each generation are embedded into two gauge multiplets $F\sim (\bm{4}, \bm{2}, \bm{1})$ and $F^c\sim (\overline{\bm{4}}, \bm{1}, \overline{\bm{2}})$. When introducing the $A_4$ modular group, the three generations of each of the gauge multiplets of matter fields can be assigned to either a triplet or three one-dimensional representations of the $A_4$ modular group. As a consequence, the different elements of the Yukawa couplings are modular forms of level 3 and they are correlated by the $A_4$. In this work, we shall perform a comprehensive analysis of the $\text{PS}\times A_4$ modular models, and all possible combinations of representation assignments of $F$ and $F^c$ under $A_4$ are considered. The goal is to determine the simplest Pati-Salam GUT models with the $A_4$ modular symmetry.

The layout of the remainder of the paper is as follows. We briefly recapitulate the formalism of the modular flavor symmetry in section~\ref{sec:modular-symmetry}. The Yukawa couplings and the fermion mass matrices in PS model are discussed in section~\ref{sec:Pati-Salam-GUT-Fmass}, and we present the general form of the Yukawa matrices invariant under the $A_4$ modular symmetry. A numerical analysis is performed and four benchmark models are presented in section~\ref{sec:numerical-results}. The predictions for the fermion masses and mixing parameters as well as the effective mass of neutrinoless double decay are studied. Finally we conclude the paper in section~\ref{sec:conclusion}. The $A_4$ modular group and the higher weight modular forms of level 3 are reprorted in the Appendix~\ref{app:A4-MDF}.

\section{\label{sec:modular-symmetry}Modular flavor symmetry }

In this section, we give a brief review on the modular flavor symmetry. The full modular group $\Gamma\cong \text{SL}(2, \mathbb{Z})$ is the group of two-dimensional matrices with integral entries and unit determinant,
\begin{equation}
\Gamma=\left\{\begin{pmatrix}
a  ~&~  b \\
c  ~&~ d
\end{pmatrix}\Big|a, b, c, d\in\mathbb{Z},~~ad-bc=1\right\}\,.
\end{equation}
The modular group has infinite elements and it can be generated by two generators $S$ and $T$,
\begin{equation}
S=\begin{pmatrix}
0  ~&~ 1\\
-1  ~&~ 0
\end{pmatrix},~~~T=\begin{pmatrix}
1 ~&~ 1\\
0 ~&~ 1
\end{pmatrix}\,,
\end{equation}
which satisfy the following relations:
\begin{equation}
\label{eq:modu-gen}S^4=(ST)^3=1,~~S^2T=TS^2\,.
\end{equation}
The modular symmetry is ubiquitous in string compactifications, and it is geometrical symmetry of the extra compact space. In the simple toroidal compactification, the two-dimensional torus $T^2$ is described as the quotient $T^{2}=\mathbb{C}/\Lambda_{\omega_1, \omega_2}$, where $\mathbb{C}$ refers to the whole complex plane $\mathbb{C}$ and $\Lambda_{\omega_1, \omega_2}=\left\{m\omega_1+n\omega_2, m,n\in\mathbb{Z}\right\}$ denotes a two-dimensional lattice with the basis vectors $\omega_1$ and $\omega_2$. The lattice is left invariant under a change in lattice basis vectors
only and if only
\begin{equation}
\begin{pmatrix}
\omega_1\\
\omega_2
\end{pmatrix}\rightarrow\begin{pmatrix}
\omega'_1\\
\omega'_2
\end{pmatrix}=\begin{pmatrix}
a  ~&~  b \\
c  ~&~ d
\end{pmatrix}\begin{pmatrix}
\omega_1\\
\omega_2
\end{pmatrix},\qquad \begin{pmatrix}
a  ~&~  b \\
c  ~&~ d
\end{pmatrix}\in\Gamma\,.
\end{equation}
The torus is characterized by the complex modulus $\tau=\omega_1/\omega_2$ up to rotation and scale transformations, without loss of generality we can limit $\tau$ to the upper half complex plane with $\text{Im}(\tau)>0$.  The two tori related by modular transformations would be identical, i.e.
\begin{equation}
\label{eq:modular-trans}\tau\xrightarrow{\gamma}\tau'=\frac{\omega'_1}{\omega'_2}=\frac{a\tau+b}{c\tau+d}\equiv\gamma\tau,~~~\text{Im}(\tau)>0,~~~\gamma=\begin{pmatrix}
a  ~&~  b \\
c  ~&~ d
\end{pmatrix}\,.
\end{equation}
Thus the action of the generators $S$ and $T$ is
\begin{equation}
\tau\xrightarrow{S}-\frac{1}{\tau},~~~\tau\xrightarrow{T}\tau+1\,.
\end{equation}
From Eq.~\eqref{eq:modular-trans} we see that $\gamma\in\Gamma$ and $-\gamma\in\Gamma$ define the same transformation of $\tau$. Employing modular transformations, one can always restrict $\tau$ to the fundamental domain defined by
\begin{eqnarray}
\label{eq:fundamental-domain}\mathcal{D}=\left\{\tau\,\Big|\,\text{Im}(\tau)>0, |\text{Re}(\tau)|\leq \frac{1}{2}, |\tau|\geq1\right\}\,.
\end{eqnarray}
Any value of $\tau$ in the upper-half plane can be mapped into the fundamental domain $\mathcal{D}$ by a suitable modular transformation, but no two points in the interior of $\mathcal{D}$ are related under the modular group. Hence the fundamental domain $\mathcal{D}$ is a representative set of the physically inequivalent modulus. Notice that the left boundary of $\mathcal{D}$ with $\text{Re}(\tau)=-1/2$ is related to the right boundary of $\text{Re}(\tau)=1/2$ by the $T$ transformation, and the $S$ transformation maps the left unit arc $\tau=e^{i\theta} (\pi/2\leq\theta\leq2\pi/3)$ on the boundary is related to the right unit arc $\tau=e^{i\theta} (\pi/3\leq\theta\leq\pi/2)$ by the $S$ transformation.

The modular symmetry provides an origin of the discrete flavor symmetry through the quotient,
\begin{equation}
\Gamma_N=\Gamma/\pm\Gamma(N),~~~\Gamma'_N=\Gamma/\Gamma(N)\,,
\end{equation}
where $\Gamma_N$ and $\Gamma'_N$ are the inhomogeneous and homogenous finite modular groups respectively, and $\Gamma(N)$ is the principal normal subgroup of level $N$,
\begin{equation}
\Gamma(N)=\left\{\begin{pmatrix}
a  ~&~  b \\
c  ~&~ d
\end{pmatrix}\in \Gamma, ~\begin{pmatrix}
a  ~&~  b \\
c  ~&~ d
\end{pmatrix}=\begin{pmatrix}
1  ~&~ 0 \\
0 ~&~ 1
\end{pmatrix}\;(\text{mod}~N)
\right\}\,,
\end{equation}
which implies $T^{N}\in\Gamma(N)$. The inhomogeneous finite modular groups $\Gamma_N$ for $N=2, 3, 4, 5$ are isomorphic to the permutation groups $S_3$, $A_4$, $S_4$ and $A_5$ respectively~\cite{deAdelhartToorop:2011re,Feruglio:2017spp}, and $\Gamma'_N$ is the double cover of $\Gamma_N$~\cite{Liu:2019khw}.

Modular invariance is the core concept of modular flavor symmetry, and the theory is assumed to be invariant under the modular transformation~\cite{Feruglio:2017spp}. In the framework of $\mathcal{N}=1$ global supersymmetry, the modulus $\tau$ is a chiral supermultiplet and its scalar component is restricted to the upper half of the complex plane, and the action can be generally written as
\begin{equation}
\mathcal{S}=\int d^4xd^2\theta d^2\bar{\theta}\, \mathcal{K}(\tau, \bar{\tau}; \Phi_I, \bar{\Phi}_I)+\left[\int d^4x d^2\theta\, \mathcal{W}(\tau, \Phi_I)+\text{h.c.}\right]\,,
\end{equation}
where the K\"ahler potential $\mathcal{K}(\tau, \bar{\tau}; \Phi_I, \bar{\Phi}_I)$ is a real gauge-invariant function of the chiral superfields $\tau$, $\Phi_I$ and their conjugates, the superpotential $\mathcal{W}(\tau, \Phi_I)$ is a holomorphic gauge invariant function of the chiral superfields $\tau$, $\Phi_I$. Under the action of $\gamma\in\Gamma$, the superfield $\Phi_I$ non-linearly transforms as
\begin{equation}
\Phi_I\xrightarrow{\gamma}(c\tau+d)^{-k_I}\rho_I(\gamma)\Phi_I\,,
\end{equation}
where the weight $k_I$ is an integer and $\rho_I$ is a unitary representation of the finite modular group $\Gamma_N$ or $\Gamma'_N$.
In the vast literature, the K\"ahler potential is assumed to be the minimal form
\begin{equation}
\mathcal{K}(\tau, \bar{\tau}; \Phi_I, \bar{\Phi}_I)=-h\log(-i\tau+i\bar{\tau})+\sum_I\frac{\bar{\Phi}_I\Phi_I}{(-i\tau+i\bar{\tau})^{k_I}}\,,
\end{equation}
which is invariant up to K\"ahler transformations. It gives the kinetic terms of for the scalar components of $\tau$ and $\Phi_I$ after the modulus get a VEV. Notice that there are many terms compatible with the modular symmetry in the K\"ahler potential so that the predictive power of modular flavor symmetry would be reduced. The K\"ahler potential is severely constrained in the paradigm of eclectic flavor group which is an extension of modular symmetry by traditional flavor symmetry, and the above minimal K\"ahler potential could be achieved~\cite{Nilles:2020nnc,Nilles:2020kgo,Nilles:2020tdp,Chen:2021prl,Ding:2023ynd,Li:2023dvm}.

As regards the superpotential $\mathcal{W}(\tau, \Phi_I)$, and it can be expanded in power series of $\Phi_I$ as follows
\begin{equation}
\mathcal{W}(\tau, \Phi_I)=\sum_n Y_{I_1\ldots I_n}(\tau)\Phi_{I_1}\ldots \Phi_{I_n}\,.
\end{equation}
Modular invariance of $\mathcal{W}$ requires that $Y_{I_1\ldots I_n}(\tau)$ should be a modular forms of weight $k_Y$ and level $N$ transforming in the representation $\rho_{Y}$ of $\Gamma_N$ (or $\Gamma'_N$), i.e.,
\begin{equation}
Y_{I_1\ldots I_n}(\tau)\xrightarrow{\gamma} Y_{I_1\ldots I_n}(\gamma\tau)=(c\tau+d)^{k_Y}\rho_{Y}(\gamma)Y_{I_1\ldots I_n}(\tau)\,.
\end{equation}
The modular weights and the representations should satisfy the conditions
\begin{equation}
k_Y=k_{I_1}+\ldots+ k_{I_n}\,,\qquad \rho_{Y}\otimes \rho_{I_1}\otimes\ldots\otimes\rho_{I_n}\supset \bm{1}\,,
\end{equation}
where $\bm{1}$ denotes the trivial singlet of $\Gamma_N$ (or $\Gamma'_N$).

In the present work, we shall concerned with the inhomogeneous finite modular group $A_4$ of level $N=3$. The even weight modular forms of level 3 can be organized into multiplets of $A_4$. The modular forms of wight $k$ at level 3 span a linear space of dimension $k+1$, the three basis vectors of the weight 2 and level 3 modular forms $Y^{(2)}_{\bm{3}}=(Y_1, Y_2, Y_3)^T$ can be chosen as
\begin{equation}
\label{eq:MF-w2l3}Y_1(\tau)=\varepsilon^2(\tau),~~~~Y_2(\tau)=\sqrt{2}\,\vartheta(\tau)\varepsilon(\tau),~~~~Y_3(\tau)=-\vartheta^2(\tau)\,,
\end{equation}
with $\varepsilon(\tau)$ and $\vartheta(\tau)$ given by
\begin{equation}
\vartheta(\tau)=3\sqrt{2}\,\frac{\eta^3(3\tau)}{\eta(\tau)},~~~~\varepsilon(\tau)=-\frac{3\eta^3(3\tau)+\eta^3(\tau/3)}{\eta(\tau)}\,.
\end{equation}
Here $\eta(\tau)$ is the Dedekind eta function defined by
\begin{equation}
\eta(\tau)=q^{1/24}\prod_{n=1}^\infty \left(1-q^n \right)\,,\quad q\equiv e^{i 2 \pi\tau}\,.
\end{equation}
It is notable that $Y_1(\tau)$, $Y_2(\tau)$ and $Y_3(\tau)$ satisfy the constraint $Y_2^2+2 Y_1 Y_3=0$ and they form an irreducible triplet of $A_4$, i.e. $Y^{(2)}_{\bm{3}}(\tau)\equiv\left(Y_1(\tau), Y_2(\tau), Y_3(\tau)\right)^{T}$. The $q$-expansion of $Y_i(\tau)$ reads as
\begin{align}
\nonumber&Y_1(\tau)=1 + 12q + 36q^2 + 12q^3 + 84q^4 + 72q^5+36q^6+96q^7+180q^8+\dots \,, \\
\nonumber&Y_2(\tau)=-6q^{1/3}\left(1 + 7q + 8q^2 + 18q^3 + 14q^4+31q^5+20q^6+36q^7+31q^8+\dots\right)\,, \\
&Y_3(\tau)=-18q^{2/3}\left(1 + 2q + 5q^2 + 4q^3 + 8q^4 +6q^5+14q^6+8q^7+14q^8+\dots\right)\,.
\end{align}
Other higher weight modular forms at level 3 can be generated from the tensor products of $Y^{(2)}_{\bm{3}}(\tau)$, as listed in Appendix~\ref{app:A4-MDF}.

\section{Fermion masses and modular invariance  in Pati-Salam GUTs }
\label{sec:Pati-Salam-GUT-Fmass}

In the present work, we intend to impose modular flavor symmetry in Pati-Salam grand unification theory (GUT) to explain the flavor structure of quarks and leptons. For illustration, we take the finite modular group $\Gamma_3\cong A_4$ which is the simplest finite group with three dimensional irreducible representation. The GUT Model is based on the Pati-Salam gauge group~\cite{Pati:1974yy},
\begin{equation}
G_{422} = SU(4)_C\times SU(2)_L\times SU(2)_R \,,
\end{equation}
which is an extension of the QCD gauge group, the PS GUT theory embeds all SM fermions of a generation with the right-handed neutrino into two chiral multiplets. The lepton is treated as the fourth color such that the strong interaction $SU(3)_C$ gauge group is extended to $SU(4)_C$, and $SU(2)_R$ is the right-handed gauge group similar to the SM $SU(2)_L$ weak interaction. In the PS GUT, the left-handed fermion multiplet is usually denoted as $F$ and the $F^c$ is defined as the CP conjugate of the right-handed fermion multiplet,
\begin{eqnarray}
\nonumber F_i  &=&  (\bm{4}, \bm{2}, \bm{1})_i =
\begin{pmatrix}
  u^r~  & ~u^g~ & ~u^b~  & ~\nu  \\
  d^r~  & ~d^g~  & ~d^b~  & ~e
\end{pmatrix}_i  \,,\\
\nonumber F^c_i  &=& (\overline{\bm{4}}, \bm{1}, \overline{\bm{2}})_i=
\begin{pmatrix}
 u^{r\,c}~  & u^{g\,c}~ & u^{b\,c}~  & \nu^c \\
d^{r\,c} ~ & d^{g\,c} ~ & d^{b\,c} ~ & e^c
\end{pmatrix}_i \,,\\
\text{or}~~F^c_i  &=& (\overline{\bm{4}}, \bm{1}, \bm{2})_i=
\begin{pmatrix}
 ~~d^{r\,c}  & ~~d^{g\,c}  & ~~d^{b\,c}  &~~e^c\\
 - u^{r\,c}~  & -u^{g\,c}~ & -u^{b\,c}~  & -\nu^c
\end{pmatrix}_i  \,.
\end{eqnarray}
The subscript $i=1,2,3$ is the family index and the superscript $r,g,b$ refer to three colors: red, green and blue in the SM $SU(3)_C$ gauge group. The fermion masses are generated by the Yukawa couplings of fermion bilinears in the spinor representation of $ (\bm{4}, \bm{2}, \bm{1})$ and $(\overline{\bm{4}}, \bm{1}, \bm{2})$ with Higgs fields multiplets. The tensor products of the fermion bilinears under the PS group are
\begin{eqnarray}
\nonumber
(\bm{4}, \bm{2}, \bm{1}) \otimes (\overline{\bm{4}}, \bm{1}, \bm{2}) &=& (\bm{1} \oplus \bm{15}, \bm{2}, \bm{2})\,, \\
\nonumber (\bm{4}, \bm{2}, \bm{1}) \otimes (\bm{4}, \bm{2}, \bm{1}) &=& (\bm{6}_{\text{A}} \oplus \bm{10}_{\text{S}}, \bm{1}_{\text{A}} \oplus \bm{3}_{\text{S}}, \bm{1})\,, \\
 (\overline{\bm{4}},\bm{1}, \bm{2}) \otimes (\overline{\bm{4}}, \bm{1}, \bm{2})  &=& (\overline{\bm{6}}_{\text{A}} \oplus \overline{\bm{10}}_{\text{S}},\bm{1},\bm{1}_{\text{A}} \oplus \bm{3}_{\text{S}})\,,
\end{eqnarray}
where the subscripts $S$ and $A$ stand for the symmetric and antisymmetric contractions of the tensor products respectively, note the $SU(4)_C$ representations $\bm{6}_{\text{A}}$ and $\overline{\bm{6}}_{\text{A}}$ are equivalent. In order to have Yukawa couplings as well as right-handed neutrino mass terms invariant under the PS gauge, we assume the existence of the following Higgs multilpets under the PS group,
\begin{eqnarray}
\label{eq:Higgs-Yukawa}\Phi = (\bm{1},\bm{2},\bm{2}) ,\quad\quad \Sigma= (\bm{15},\bm{2},\bm{2})\,,\quad\quad \Delta_R  = (\bm{10},\bm{1},\bm{3})\,.
\end{eqnarray}
In this work, we shall formulate our model in the framework of the renormalizable Supersymmetric (SUSY) PS GUT for simplicity, thus the most general form of the Yukawa superpotential in renormalizable PS models reads as follow,
\begin{equation}
\label{eq:WY}
\mathcal{W}_Y=\mathcal{Y}^{1}_{ij}F^{c}_{i} F_{j}\Phi+\mathcal{Y}^{15}_{ij}F^{c}_{i} F_{j}\Sigma +\mathcal{Y}^{10_R}_{ij}F^{c}_{i}F^{c}_{j}\Delta_{R}\,,
\end{equation}
where $i,j=1, 2, 3$ are indices of generation, the Yukawa coupling matrices $\mathcal{Y}^{1}$ and $\mathcal{Y}^{15}$ are generic $3\times3$ complex matrix in flavor spaces and their matrix elements are not constrained by the PS gauge symmetry. It is notable that $\mathcal{Y}^{10_R}$ is a symmetric three dimensional matrix, i.e.
\begin{equation}
\mathcal{Y}^{10_R}_{ij}=\mathcal{Y}^{10_R}_{ji}\,.
\end{equation}
Besides the Higgs multiplets shown in Eq.~\eqref{eq:Higgs-Yukawa}, the breaking of the PS symmetry requires the following minimal set of Higgs-like supermultiplets: $A=(\mathbf{15}, \mathbf{1}, \mathbf{1})$, $\overline{\Delta}_R=(\overline{\bm{10}}, \bm{1}, \bm{3})$, $\Delta_L=(\bm{\overline{10}},\bm{3},\bm{1})$ and $\overline{\Delta}_L=(\bm{10},\bm{3},\bm{1})$~\cite{Melfo:2003xi}. The VEV of $A$ breaks $SU(4)_C$ down to $SU(3)_C\times U(1)_{B-L}$. Subsequently the VEVs of $\Delta_R$ and $\overline{\Delta}_R$ breaks $SU(2)_R\times U(1)_{B-L}$ down to the SM hypercharge $U(1)_Y$. Thus the PS group is broken into the SM gauge symmetry $SU(3)_C\times SU(2)_L\times U(1)_Y$ through the above two-step breaking. Finally the VEVs of $\Phi$ and $\Sigma$ breaks the SM into $SU(3)_C\times U(1)_{EM}$ at low energy. As a consequence, the VEVs of the Higgs multiplets have the following hierarchical pattern
\begin{equation}\label{Hhier}
\langle A\rangle> \langle\Delta_R\rangle\gg  \langle\Phi\rangle \sim \langle\Sigma\rangle \,.
\end{equation}
Notice that the fields $\Delta_L$ and $\overline{\Delta}_L$ don't develop VEVs at tree level in the usual minimal PS model, unless higher order terms are considered or other new supermultiplets such as $\Omega=(\bm{1}, \bm{3}, \bm{3})$ or $X=(\overline{\bm{10}}, \bm{2}, \bm{2})$, $\overline{X}=(\bm{10}, \bm{2}, \bm{2})$ are included further~\cite{Melfo:2003xi}. Hence the type-II seesaw contributions to the effective neutrino masses are almost
negligible in the minimal PS GUTs although the couplings $\mathcal{Y}^{10_L}_{ij}F_{i}F_{j}\Delta_{L}$ are compatible with the PS gauge symmetry. As a consequence, we shall focus on the type-I seesaw mechanism to generate neutrino masses in this work.

Under the SM group, the corresponding left-handed and right-handed fermions representation in the Pati-Salam group  have the following decomposition:
\begin{eqnarray}
\nonumber {\bf PS} \quad& {\bf \rightarrow} &\quad {\bf SU(3)_C\times SU(2)_L\times U(1)_Y}\\
\nonumber F=(\mathbf{4}, \mathbf{2}, \mathbf{1}) \quad& {\bf \rightarrow} &\quad (\mathbf{3}, \mathbf{2}, 1/6 )_{Q} +(\mathbf{1}, \mathbf{2}, -1/2)_{L}\\
F^c=(\mathbf{\overline{4}}, \mathbf{1}, \mathbf{2})  \quad& {\bf \rightarrow} &\quad (\mathbf{\overline{3}}, \mathbf{1}, -2/3)_{u^c} + (\mathbf{\overline{3}}, \mathbf{1}, 1/3 )_{d^c} +(\mathbf{1}, \mathbf{1}, 1)_{e^c} + (\mathbf{1}, \mathbf{1}, 0)_{\nu^c}
\end{eqnarray}
where the subscripts $Q$ and $L$ are the left-handed quark field and left-handed lepton field respectively. The hypercharge is embedded as
\begin{equation}\label{hypercharge}
 Y=\frac{B-L}{2}+T_{R3}\,,
\end{equation}
where $B - L$ is the (suggestively named) unbroken generator of $SU(4)_C$ which does not belong to $SU(3)_C$ and $T_{R3}$ is the third generator of $SU(2)_R$. Similarly, the decomposition of PS Higgs multiplets under the SM gauge symmetry are given by:
\begin{eqnarray}
\nonumber {\bf PS} \quad& {\bf \rightarrow} &\quad {\bf SU(3)_C\times SU(2)_L\times U(1)_Y}\\
\Phi=(\mathbf{1}, \mathbf{2}, \mathbf{2})&\rightarrow&(\mathbf{1}, \mathbf{2}, 1/2)+(\mathbf{1}, \mathbf{2}, -1/2)\,,\\
\nonumber \Sigma=(\mathbf{15}, \mathbf{2}, \mathbf{2})&\rightarrow&(\mathbf{1}, \mathbf{2}, 1/2) + (\mathbf{1}, \mathbf{2}, -1/2) + (\mathbf{3}, \mathbf{2}, 1/6) + (\overline{\mathbf{3}}, \mathbf{2}, -1/6) + (\mathbf{3}, \mathbf{2}, 7/6) \\
&& + (\mathbf{\overline{3}}, \mathbf{2}, -7/6) + (\mathbf{8}, \mathbf{2}, 1/2) + (\mathbf{8}, \mathbf{2}, -1/2)\,, \\
\nonumber \Delta_R=(\mathbf{10}, \mathbf{1}, \mathbf{3})&\rightarrow& (\mathbf{1}, \mathbf{1}, 0) + (\mathbf{1}, \mathbf{1}, -1) +(\mathbf{1}, \mathbf{1}, -2) + (\mathbf{3}, \mathbf{1}, 2/3) + (\mathbf{3}, \mathbf{1}, -1/3) \\
&& + (\mathbf{3}, \mathbf{1}, -4/3) +(\mathbf{6}, \mathbf{1}, 4/3) + (\mathbf{6}, \mathbf{1}, -1/3) + (\mathbf{6}, \mathbf{1}, -2/3)\,,
\end{eqnarray}
We see that the PS Higgs fields $\Phi$, $\Sigma$ and $\Delta_R$ have the minimal supersymmetric standard model (MSSM) Higgs components, i.e.
\begin{eqnarray}
\nonumber&& \Phi \supset(\bm{1}, \bm{2}, 1/2)\oplus(\bm{1}, \bm{2}, -1/2)\equiv \Phi_u\oplus\Phi_d\,,\\
\nonumber&&\Sigma\supset(\bm{1}, \bm{2}, 1/2)\oplus(\bm{1}, \bm{2}, -1/2)\equiv \Sigma_u\oplus\Sigma_d\,,\\
&&\Delta_R\supset(\bm{1}, \bm{1}, 0) \,,
\end{eqnarray}
where $\Phi_u$ and $\Sigma_u$ have the same SM quantum numbers as the up-type Higgs field $H_u$, $\Phi_d$ and $\Sigma_d$ transform in the same way as the down-type Higgs field $H_d$. Decomposing the Yukawa superpotential in Eq.~\eqref{eq:WY}, we find the quark and lepton mass matrices are of the following form after breaking to $SU(3)_C\times U(1)_{EM}$ symmetry,
\begin{eqnarray}
\nonumber & M_u =\mathcal{Y}^{1} v^u_\Phi +\mathcal{Y}^{15}v^u_\Sigma\,,\quad M_d =\mathcal{Y}^{1} v^d_\Phi + \mathcal{Y}^{15}v^d_\Sigma \,,\\
\label{eq:mass-matrices-1}&M_D =  \mathcal{Y}^{1} v^u_\Phi-3\mathcal{Y}^{15}v^u_\Sigma\,,\quad M_e = \mathcal{Y}^{1} v^d_\Phi - 3 \mathcal{Y}^{15}v^d_\Sigma \,,\quad M_R =  \mathcal{Y}^{10_R} v_R\,,
\end{eqnarray}
where the VEVs are defined as
\begin{eqnarray}
\langle\Delta_R\rangle = v_R,\quad \langle\Phi_u\rangle = v^u_\Phi,\quad
 \langle\Phi_d\rangle =  v^d_\Phi,\quad \langle\Sigma_u\rangle = v^u_\Sigma,\quad \langle\Sigma_d\rangle = v^d_\Sigma\,,
\end{eqnarray}
where $M_u$, $M_d$ are the up-type and down-type quark mass matrices, $M_e$ is the charged lepton mass matrix, $M_D$ is the neutrino Dirac mass matrix and $M_R$ is the right-handed Majorana neutrino mass matrix. The Georgi-Jarlskog factor $-3$ in Eq.~\eqref{eq:mass-matrices-1} can reconcile the different mass spectrum of down-type quarks and charged leptons. The light neutrino mass matrix is given by the famous  type I seesaw formula
\begin{equation}
\label{seesaw}
M_{\nu}=- M^T_D  M_R^{-1} M_D\,.
\end{equation}
It is convenient to redefine the Yukawa coupling parameters as follows
\begin{eqnarray}
\nonumber   \mathcal{\widetilde{Y}}^{1}   &=& \frac{v^u_\Phi}{v_u}\mathcal{Y}^{1},\quad\quad  \mathcal{\widetilde{Y}}^{15}  = \frac{v^d_\Sigma}{v_u}\frac{v^u_\Phi }{v^d_\Phi}\mathcal{Y}^{15}\,, \\
 r_1&=&\frac{v^d_\Phi}{v^u_\Phi}\frac{v_u}{v_d},\quad\quad r_2=\frac{v^u_\Sigma}{v^d_\Sigma}\frac{v^d_\Phi}{v^u_\Phi}\,,
\end{eqnarray}
where $v_u$ and $v_d$ are the VEVs of the MSSM Higgs pair $H_u$ and $H_d$. The parameters $r_a$\,$(a=1,2)$ are mixing parameters which relate $H_{u,d}$ to the $SU(2)_L$ doublet components of the PS Higgs multiplets. Notice that $\mathcal{\widetilde{Y}}^{1}$ and $\mathcal{\widetilde{Y}}^{15}$ are proportional to the Yukawa matrices $\mathcal{Y}^{1}$ and $\mathcal{Y}^{15}$ respectively, and the coefficients $\frac{v^u_\Phi}{v_u}$ and $\frac{v^d_\Sigma}{v_u}\frac{v^u_\Phi }{v^d_\Phi}$ can be absorbed into the coupling constants. Then the mass matrices of the quarks and leptons can be written as
\begin{eqnarray}
\nonumber &M_u  = \left(\mathcal{\widetilde{Y}}^{1} + r_2\mathcal{\widetilde{Y}}^{15}\right)v_u\,,\quad M_d =r_1\left(\mathcal{\widetilde{Y}}^{1} + \mathcal{\widetilde{Y}}^{15}\right)v_d \,,\\
\label{eq:mass-matrices-final}&M_D =  \left(\mathcal{\widetilde{Y}}^{1} -3 r_2\mathcal{\widetilde{Y}}^{15}\right)v_u\,,\quad M_e =r_1\left(\mathcal{\widetilde{Y}}^{1} -3 \mathcal{\widetilde{Y}}^{15}\right)v_d \,, \quad M_R =  \mathcal{Y}^{10_R} v_R\,.
\end{eqnarray}
The effective neutrino mass matrix is still given by Eq.~\eqref{seesaw}.

\subsection{Combining Pati-Salam group with $A_4$ modular symmetry}

In PS GUT, the renormalizable Yukawa superpotential compatible with modular symmetry can be written as
\begin{eqnarray}\label{eq:WY-Mod-sym}
\nonumber\mathcal{W}_Y &=& \sum_{\{\bm{r}_a,\bm{r}'_a\}}\alpha_a\left(\left(F^{c}F\Phi\right)_{\bm{r}'_a}Y_{\bm{r}_a}^{(k_F+k_{F^c}+k_{\Phi})}(\tau)\right)_{\bm{1}}+
\sum_{\{\bm{r}_b,\bm{r}'_b\}}\beta_b\left(\left(F^{c}F\Sigma\right)_{\bm{r}'_b}Y_{\bm{r}_b}^{(k_F+k_{F^c}+k_{\Sigma})}(\tau)\right)_{\bm{1}}\\
&&+\sum_{\{\bm{r}_d,\bm{r}'_d\}}\gamma_d\left(\left(F^{c}F^{c}\Delta_{R}\right)_{\bm{r}'_d}Y_{\bm{r}_d}^{(2k_{F^c}+k_{\Delta_R})}(\tau)\right)_{\bm{1}}\,,
\end{eqnarray}
where the $A_4$ representations $\bm{r}_{a,b,d}$ and $\bm{r}'_{a,b,d}$ satisfy $\bm{r}'_a\otimes \bm{r}_a=\bm{r}'_b\otimes \bm{r}_b=\bm{r}'_d\otimes \bm{r}_d=\bm{1}$. The expressions of the modular forms $Y_{\bm{r}_a}^{(k_F+k_{F^c}+k_{\Phi})}(\tau)$, $Y_{\bm{r}_b}^{(k_F+k_{F^c}+k_{\Sigma})}(\tau)$ and $Y_{\bm{r}_d}^{(2k_{F^c}+k_{\Delta_R})}(\tau)$ can be read from Eqs.~(\ref{eq:MF-w2l3}, \ref{eq:MF-w4l3}, \ref{eq:MF-w6l3}, \ref{eq:MF-w8l3}, \ref{eq:MF-w10l3}). Notice that one has to sum over all possible contractions and the contributions of all independent modular multiplets at the relevant weights. In comparison with the PS models based on traditional flavor symmetry~\cite{deAdelhartToorop:2010vtu,King:2013hoa,King:2014iia,CarcamoHernandez:2017owh}, the flavon fields are replaced by the modular forms, the complication of vacuum alignment is eliminated so that the models become much simpler.

Without loss of generality, we can set the modular weights of the Higgs fields $\Phi$ and $\Delta_R$ to be vanishing, i.e. $k_{\Phi}=k_{\Delta_{R}}=0$. In the following, we assume that all the PS Higgs multiplets are singlets of the $A_4$ modular symmetry, and we shall give the Yukawa matrices for different $A_4$ transformation properties of $F$ and $F^c$ by using the multiplication rules in Eq.~\eqref{eq:tensor-product-A4}.

\begin{itemize}

\item{$F^c\sim {\mathbf3}$,\quad $F\sim {\mathbf3}$}

In this case, both left-handed and right-handed fermions transform as irreducible triplets under $A_4$, and one can read off the Yukawa coupling $\mathcal{Y}^{1}$ as follow,
\begin{eqnarray}
\nonumber\mathcal{Y}^{1}&=&\alpha_1Y_{\mathbf1}^{(k)}(\tau)\left(
\begin{matrix}
 1 ~& 0 ~& 0 \\
 0 ~& 0 ~& 1 \\
 0 ~& 1 ~& 0 \\
\end{matrix}
\right)+\alpha_2Y_{\mathbf1'}^{(k)}(\tau)\left(
\begin{matrix}
 0 ~& 0 ~& 1 \\
 0 ~& 1 ~& 0 \\
 1 ~& 0 ~& 0 \\
\end{matrix}
\right)\\
\nonumber &&\qquad +\alpha_3Y_{\mathbf1''}^{(k)}(\tau)\left(
\begin{matrix}
 0 ~& 1 ~& 0 \\
 1 ~& 0 ~& 0 \\
 0 ~& 0 ~& 1 \\
\end{matrix}
\right)+\alpha_S\left(\begin{matrix}
 2Y^{(k)}_{{{\mathbf3}},1}(\tau) ~&~ -Y^{(k)}_{{{\mathbf3}},3}(\tau)  ~&~ -Y^{(k)}_{{{\mathbf3}},2}(\tau) \\
 -Y^{(k)}_{{{\mathbf3}},3}(\tau) ~&~ 2Y^{(k)}_{{{\mathbf3}},2}(\tau)  ~&~ -Y^{(k)}_{{{\mathbf3}},1}(\tau) \\
 -Y^{(k)}_{{{\mathbf3}},2}(\tau) ~&~ -Y^{(k)}_{{{\mathbf3}},1}(\tau)  ~&~ 2Y^{(k)}_{{{\mathbf3}},3}(\tau)
\end{matrix}
\right)\\
&&\qquad +\alpha_A\left(\begin{matrix}
 0 ~&~ -Y^{(k)}_{{{\mathbf3}},3}(\tau)  ~&~ Y^{(k)}_{{{\mathbf3}},2}(\tau) \\
 Y^{(k)}_{{{\mathbf3}},3}(\tau) ~&~ 0  ~&~ -Y^{(k)}_{{{\mathbf3}},1}(\tau) \\
 -Y^{(k)}_{{{\mathbf3}},2}(\tau) ~&~ Y^{(k)}_{{{\mathbf3}},1}(\tau)  ~&~ 0
\end{matrix}
\right)\,,
\end{eqnarray}
with $k=k_F+k_{F^c}$ and $\mathcal{Y}^{15}$ takes similar form just by replacing $\alpha_i$ with $\beta_i$. Since the Yukawa matrix $\mathcal{Y}^{10_R}$ is symmetric, the antisymmetric triplet contraction have no contribution, thus it can be written as
\begin{eqnarray}\label{eq:tripletYL}
\nonumber&&\mathcal{Y}^{10_R}\Big|_{k=2k_{F}}=\gamma_1Y_{\mathbf1}^{(k)}(\tau)\left(
\begin{matrix}
 1 ~& 0 ~& 0 \\
 0 ~& 0 ~& 1 \\
 0 ~& 1 ~& 0 \\
\end{matrix}
\right)+\gamma_2Y_{\mathbf1'}^{(k)}(\tau)\left(
\begin{matrix}
 0 ~& 0 ~& 1 \\
 0 ~& 1 ~& 0 \\
 1 ~& 0 ~& 0 \\
\end{matrix}
\right)\\
&&\qquad +\gamma_3Y_{\mathbf1''}^{(k)}(\tau)\left(
\begin{matrix}
 0 ~& 1 ~& 0 \\
 1 ~& 0 ~& 0 \\
 0 ~& 0 ~& 1 \\
\end{matrix}
\right)+\gamma_4\left(\begin{matrix}
 2Y^{(k)}_{{{\mathbf3}},1}(\tau) ~&~ -Y^{(k)}_{{{\mathbf3}},3}(\tau)  ~&~ -Y^{(k)}_{{{\mathbf3}},2}(\tau) \\
 -Y^{(k)}_{{{\mathbf3}},3}(\tau) ~&~ 2Y^{(k)}_{{{\mathbf3}},2}(\tau)  ~&~ -Y^{(k)}_{{{\mathbf3}},1}(\tau) \\
 -Y^{(k)}_{{{\mathbf3}},2}(\tau) ~&~ -Y^{(k)}_{{{\mathbf3}},1}(\tau)  ~&~ 2Y^{(k)}_{{{\mathbf3}},3}(\tau)
\end{matrix}
\right)\,.
\end{eqnarray}

\item{$F^c\sim {\mathbf3} $,\quad $F\sim {\mathbf1^{i_1}}\oplus{\mathbf1^{i_2}}\oplus{\mathbf1^{i_3}}$}

In this case,  we denote the $A_4$ singlet representation as ${\mathbf1}\equiv{\mathbf1^{0}}$, ${\mathbf1^{'}}\equiv{\mathbf1^{1}}$, ${\mathbf1^{''}}\equiv{\mathbf1^{2}}$ for notational simplicity with $i_1, i_2, i_3=0, 1, 2$. Then the $A_4$ modular symmetry constrains the Yukawa couplings $\mathcal{Y}^{1}$ and $\mathcal{Y}^{15}$ to be
\begin{eqnarray}
\nonumber \mathcal{Y}^{1}&=&\left(\alpha_1 C^{(k_1)}_{i_1}(\tau),\,\alpha_2 C^{(k_2)}_{i_2}(\tau),\,\alpha_3 C^{(k_3)}_{i_3}(\tau)\right)\,, \\
\label{eq:Y1-Y15-Fc3-F1}\mathcal{Y}^{15}&=&\left(\beta_1 C^{(k_1)}_{i_1}(\tau),\,\beta_2 C^{(k_2)}_{i_2}(\tau),\,\beta_3 C^{(k_3)}_{i_3}(\tau)\right)\,,
\end{eqnarray}
where $k_i=k_{F_i}+k_{F^c}$, and $C^{(k_l)}_{i_l}$ with $l=1,2,3$ are column vectors with the following form
\begin{equation}
C^{(k)}_{0}(\tau)\equiv\begin{pmatrix}
Y^{(k)}_{{{\mathbf3}},1}(\tau) \\
Y^{(k)}_{{{\mathbf3}},3}(\tau) \\
 Y^{(k)}_{{{\mathbf3}},2}(\tau)
 \end{pmatrix},\,
C^{(k)}_{1}(\tau)\equiv\begin{pmatrix}
Y^{(k)}_{{{\mathbf3}},3}(\tau) \\
Y^{(k)}_{{{\mathbf3}},2}(\tau) \\
Y^{(k)}_{{{\mathbf3}},1}(\tau)
\end{pmatrix},\,
C^{(k)}_{2}(\tau)\equiv\begin{pmatrix}
Y^{(k)}_{{{\mathbf3}},2}(\tau) \\
Y^{(k)}_{{{\mathbf3}},1}(\tau) \\
Y^{(k)}_{{{\mathbf3}},3}(\tau)
\end{pmatrix}\,.
\end{equation}
The symmetric Yukawa matrix $\mathcal{Y}^{10_R}$ has the same form as that of Eq.~\eqref{eq:tripletYL}.

\item{ $F^c\sim {\mathbf1^{i_1}}\oplus{\mathbf1^{i_2}}\oplus{\mathbf1^{i_3}}$,\quad$F\sim {\mathbf3} $}

In comparison with previous case, the transformations of $F$ and $F^c$ under $A_4$ are interchanged. As a consequence, the Yukawa matrices  $\mathcal{Y}^{1}$ and $\mathcal{Y}^{15}$ become into their transpose, i.e.
\begin{eqnarray}
\mathcal{Y}^{1}=
\begin{pmatrix}
 \alpha_1 C^{(k_1)T}_{i_1}(\tau)\\[0.06in]
 \alpha_2 C^{(k_2)T}_{i_2}(\tau) \\[0.06in]
 \alpha_3 C^{(k_3)T}_{i_3}(\tau)
\end{pmatrix}\,,\qquad \mathcal{Y}^{15}=
\begin{pmatrix}
 \beta_1 C^{(k_1)T}_{i_1}(\tau)\\[0.06in]
 \beta_2 C^{(k_2)T}_{i_2}(\tau) \\[0.06in]
 \beta_3 C^{(k_3)T}_{i_3}(\tau)
\end{pmatrix}\,,
\end{eqnarray}
where $k_i=k_{F^c_i}+k_F$. Since $F^c$ are $A_4$ singlets,  the different entries of $\mathcal{Y}^{10_R}$ have no correlation and it is given by,
\begin{eqnarray}
 \mathcal{Y}^{10_R}\Big|_{k=2k_{F^c}} &=& \left(\begin{matrix}
  \xi_1 Y^{(2k_{F^c_1})}_{\mathbf1^{<i_1+i_1>}}(\tau) ~&~\xi_2 Y^{(k_{F^c_1}+k_{F^c_2})}_{\mathbf1^{<i_1+i_2>}}(\tau)  ~&~ \xi_3 Y^{(k_{F^c_1}+k_{F^c_3})}_{\mathbf1^{<i_1+i_3>}}(\tau) \\
  \\
  \xi_2 Y^{(k_{F^c_1}+k_{F^c_2})}_{\mathbf1^{<i_1+i_2>}}(\tau) ~&~\xi_4 Y^{(2k_{F^c_2})}_{\mathbf1^{<i_2+i_2>}}(\tau)  ~&~ \xi_5 Y^{(k_{F^c_2}+k_{F^c_3})}_{\mathbf1^{<i_2+i_3>}}(\tau) \\
  \\
 \xi_3 Y^{(k_{F^c_1}+k_{F^c_3})}_{\mathbf1^{<i_1+i_3>}}(\tau) ~&~\xi_5 Y^{(k_{F^c_2}+k_{F^c_3})}_{\mathbf1^{<i_2+i_3>}}(\tau)  ~&~ \xi_6 Y^{(2k_{F^c_3})}_{\mathbf1^{<i_3+i_3>}}(\tau)
\end{matrix}
\right)\,,
\end{eqnarray}
where we have defined the notation $<i>=i\,(\text{mod}\,\,3)$ which take values in the range of $\{0,1,2\}$.

\item{$F^c\sim {\mathbf1^{i_1}}\oplus{\mathbf1^{i_2}}\oplus{\mathbf1^{i_3}}$,\quad $F\sim {\mathbf1^{j_1}}\oplus{\mathbf1^{j_2}}\oplus{\mathbf1^{j_3}}$}

Since the modular generator $S=1$ in all singlet representations $\mathbf{1}$, $\mathbf{1}'$ and $\mathbf{1}''$, the flavor symmetry is essentially the abelian subgroup $Z_3$ generated by the modular generator $T$. The entries of the Yukawa couplings would be not correlated so that more coupling constants would be involved, therefore we do not consider this assignment.
\end{itemize}

\section{\label{sec:numerical-results}Numerical analysis}

In this section, we shall perform a comprehensive numerical analysis of the predictions for the possible PS models classified according to the $A_4$ modular transformations of the matter fields $F$ and $F^c$, the light neutrino masses are assumed to be generated by the type I seesaw mechanism, and the level 3 modular forms up to weight 10 have been considered. The modular symmetry is uniquely broken by the value of the complex modulus $\tau$. We shall treat the VEV of $\tau$ as a free complex parameter to adjust the agreement with the experimental data, and it is sufficient to restrict $\tau$ in the fundamental domain $\mathcal{D}$ of Eq.~\eqref{eq:fundamental-domain}. For any given values of the complex modulus $\tau$ and the coupling constants, one can numerically diagonalize the mass matrices of quarks and leptons and subsequently extract the values of fermion masses, mixing angles and CP violation phases.

In order to quantitatively measure whether a $\text{PS}\times A_4$ model can accommodate the experimental data of the fermion masses and flavor mixing, we perform a conventional $\chi^2$ analysis to estimate the goodness of fit, and the $\chi^{2}$ function is defined as
\begin{equation}
\chi^2=\sum_{i}\left(\frac{P_i(x)-\mu_i}{\sigma_i}\right)^2\,,
\end{equation}
where $\mu_i$ and $\sigma_i$ denote the central values and the $1\sigma$ deviations of the experimental data for flavor observables including the mass ratios, mixing angles and CP violation phases of quarks and leptons,  $P_i(x)$ are the theoretical predictions for the corresponding observables as complex nonlinear functions of the model free parameters generally denoted by $x$ here. The central values and the $1\sigma$ uncertainties of the quark masses, three quark mixing angles and one CP violation phase in the CKM mixing matrix at the GUT scale are adopted from~\cite{Ross:2007az} for $\tan\beta=10$, as listed in table~\ref{tab:parameter-values-1sigma}. As regards the observables in the lepton sector, the charged lepton masses are taken from~\cite{Ross:2007az}, we adopt the values of the neutrino mass squared differences $\Delta m_{21}^{2}$ and $\Delta m_{31}^{2}$ as well as the values of lepton mixing angles and CP phase from the latest global fit NuFIT 5.2~\cite{Esteban:2020cvm}. The renormalization group running effect on the neutrino parameters are small enough to be negligible for normal ordering neutrino masses and $\tan\beta\leq10$. It is notable that the results of the atmospheric angle $\theta^{\ell}_{23}$ depends on whether the data of Super-Kamiokande (SK) is included or not in the global analysis. The central value of $\theta^{\ell}_{23}$ is in the first (second) octant if the SK data is (not) included. However, the $3\sigma$ regions of neutrino mixing parameters weakly depend on the SK data. In the present work, we shall consider both cases with and without SK data.

\begin{table}[t!]
\centering
\begin{tabular}{|c|c|c|c|c|} \hline  \hline
Parameters & $\mu_i\pm1\sigma$ & Parameters & $\mu_i\pm1\sigma$ & $3\sigma$ regions \\ \hline
$m_{t}/\text{GeV}$ & $83.155 \pm 3.465$ & $m_{\mu}/m_{\tau}$ & $0.059 \pm 0.002$ & --\\
  $m_{u}/m_{c}$ & $0.0027 \pm 0.0006$ & $\Delta m_{21}^{2}/(10^{-5}\text{eV}^{2})$ & $7.41_{-0.20}^{+0.21}$ &$[6.82, 8.03]$  \\
  $m_{c}/m_{t}$ & $0.0025 \pm 0.0002$ & $\Delta m_{31}^{2}/(10^{-3}\text{eV}^{2})$ (WSK) & $2.507_{-0.027}^{+0.026}$ &$[2.427, 2.590]$  \\
  $m_{b}/\text{GeV}$ & $0.884 \pm 0.035$ & $\Delta m_{31}^{2}/(10^{-3}\text{eV}^{2})$ (WOSK) & $2.511_{-0.027}^{+0.028}$ &$[2.428, 2.597]$  \\
  $m_{d}/m_{s}$ & $0.051 \pm 0.007$ & $\sin^{2}\theta_{12}^{\ell}$ & $0.303\pm 0.012$ & $[0.270, 0.341]$\\
  $m_{s}/m_{b}$ & $0.019 \pm 0.002$ & $\sin^{2}\theta_{23}^{\ell}$ (WSK)& $0.451_{-0.016}^{+0.019}$ & $[0.408, 0.603]$\\
  $\theta_{12}^{q}$ & $0.229 \pm 0.001$ & $\sin^{2}\theta_{23}^{\ell}$ (WOSK)& $0.572_{-0.023}^{+0.018}$ & $[0.406, 0.620]$\\
 $\theta_{13}^{q}$ & $0.0037 \pm 0.0004$ & $\sin^{2}\theta_{13}^{\ell}$ (WSK)& $0.02225_{-0.00059}^{+0.00056}$ & $[0.02052, 0.02398]$\\
  $\theta_{23}^{q}$ & $0.0397 \pm 0.0011 $ & $\sin^{2}\theta_{13}^{\ell}$ (WOSK)& $0.02203_{-0.00059}^{+0.00056}$ & $[0.02029, 0.02391]$\\
  $\delta_{CP}^{q}/^{\circ}$ & $56.34 \pm 7.89 $ & $\delta_{CP}^{\ell}/^{\circ}$ (WSK)& $232_{-26}^{+36}$ & $[144, 350]$\\
  $m_{\tau}/\text{GeV}$ & $1.213\pm 0.052$ & $\delta_{CP}^{\ell}/^{\circ}$ (WOSK)& $197_{-25}^{+42}$ & $[108, 404]$\\
  $m_{e}/m_{\mu}$ & $0.0048 \pm 0.0002$ & & & \\
\hline \hline
\end{tabular}
\caption{\label{tab:parameter-values-1sigma} The numerical values of the fermion masses and mixing parameters at the GUT scale derived in the framework of the minimal SUSY breaking scenarios with $\tan\beta=10$ and the SUSY breaking scale $M_{\text{SUSY}}=500$ GeV~\cite{Ross:2007az}. The values of lepton mixing angles $\theta_{12, 13, 23}^{\ell}$, lepton CP violation phases $\delta^{\ell}_{CP}$ and the neutrino mass squared differences $\Delta m_{21}^{2}$, $\Delta m_{31}^{2}$ are taken from NuFIT 5.2~\cite{Esteban:2020cvm} for normal ordering of neutrino masses with/without Super-Kamiokande (WSK/WOSK) atmospheric data.}
\end{table}

We divide the total $\chi^2$ into three parts $\chi^2=\chi^2_{\ell}+\chi^2_{q}+\chi^2_{b\tau}$, where $\chi^2_{\ell}$, $\chi^2_{q}$ and $\chi^2_{b\tau}$ quantitatively characterizes how large the lepton obervables, quark observables and $m_{b}/m_{\tau}$  deviate from their experimentally central values respectively. The leptonic $\chi^2_{\ell}$ function is constructed with the data of $m_{e}/m_{\mu}$, $m_{\mu}/m_{\tau}$, $\Delta m^2_{21}/\Delta m^2_{31}$, $\sin^{2}\theta_{12}^{\ell}$, $\sin^{2}\theta_{13}^{\ell}$, $\sin^{2}\theta_{23}^{\ell}$ and $\delta_{CP}^{\ell}$. The solar neutrino mass squared difference $\Delta m^2_{21}$ can be reproduced exactly by adjusting the overall scale of the light neutrino mass matrix, therefore the $\chi^2_{\ell}$ is built with the ratio $\Delta m^2_{21}/\Delta m^2_{31}$ rather than $\Delta m^2_{21}$ and $\Delta m^2_{31}$. Analogously the quark $\chi^2_{q}$ function is constructed with the mass ratios $m_u/m_c$, $m_c/m_t$, $m_d/m_s$, $m_s/m_b$ and quark mixing parameters $\theta^q_{12}$, $\theta^q_{13}$, $\theta^q_{23}$, $\delta^q_{CP}$. The top quark mass $m_t$ and bottom quark mass $m_b$ are used to fix the overall scales of the up-type and down-type quark mass matrices. From Eq.~\eqref{eq:mass-matrices-final}, we see that the mass matrices of the down-type quarks and charged leptons depend on the same set of parameters and share a common overall scale factor $r_1v_d$, they are different solely because of the Georgi-Jarlskog factor $-3$. The measured bottom quark mass $m_b$ can be obtained by tuning this overall scale, then the tau mass $m_{\tau}$ would be a prediction, and the $\chi^2_{b\tau}$ function built with $m_b/m_{\tau}$ allows to see the compatibility with the measured tau mass.

For each possible model, the coupling constants are freely varied with the absolute values in the range of $[0, 10^5]$ and the phases in the range of $[0, 2\pi]$, we numerically minimize the $\chi^2$ function to find out the best fit point and the corresponding predictions for the fermion masses and mixing parameters. We find that many $\text{PS}\times A_4$ models are compatible with the experimental data, and the phenomenologically viable models depend on at least 19 real parameters including the real and imaginary parts of the modulus $\tau$. We notice that the models would be strongly constrained by modular symmetry and a large number of parameters are required to accommodate the data, if both $F_i$ and $F^c_i$ transform as triplet under $A_4$. It is too lengthy to list all the viable models with the minimum number of parameters, we shall present four benchmark models in which either $F_i$ or $F^c_i$ is a $A_4$ triplet. All the three PS Higgs multiplets $\Phi$, $\Sigma$ and $\Delta_R$ are assumed to be invariant singlet of $A_4$. By shifting the modular weights of $F_i$ and $F^c_i$, one can set the modular weights of $\Phi$ and $\Delta_R$ to be vanishing with $k_{\Phi}=k_{\Delta_R}=0$. As a consequence, each $\text{PS}\times A_4$ invariant model is fully specified by the representations and modular weights of the matter fields $F_i$ and $F^c_i$ as well as the modular weight $k_{\Sigma}$ of the Higgs field $\Sigma$.

\subsection{$F^c\sim\mathbf{3}$}

The CP conjugate of the right-handed fermions $F^c$ transform as a triplet $\mathbf{3}$ under the $A_4$ modular symmetry, and the left-handed fermions $F_i$ are assigned to be $A_4$ singlets. As a consequence, the Yukawa couplings of $\Phi$ and $\Sigma$ require the triplet modular forms of level 3. In this subsection, we give two typical $\text{PS}$ models based on $A_4$ modular symmetry which can accommodate all the experimental data of quarks and leptons with 19 free parameters.

\begin{figure}[hptb!]
\centering
\includegraphics[width=6.5in]{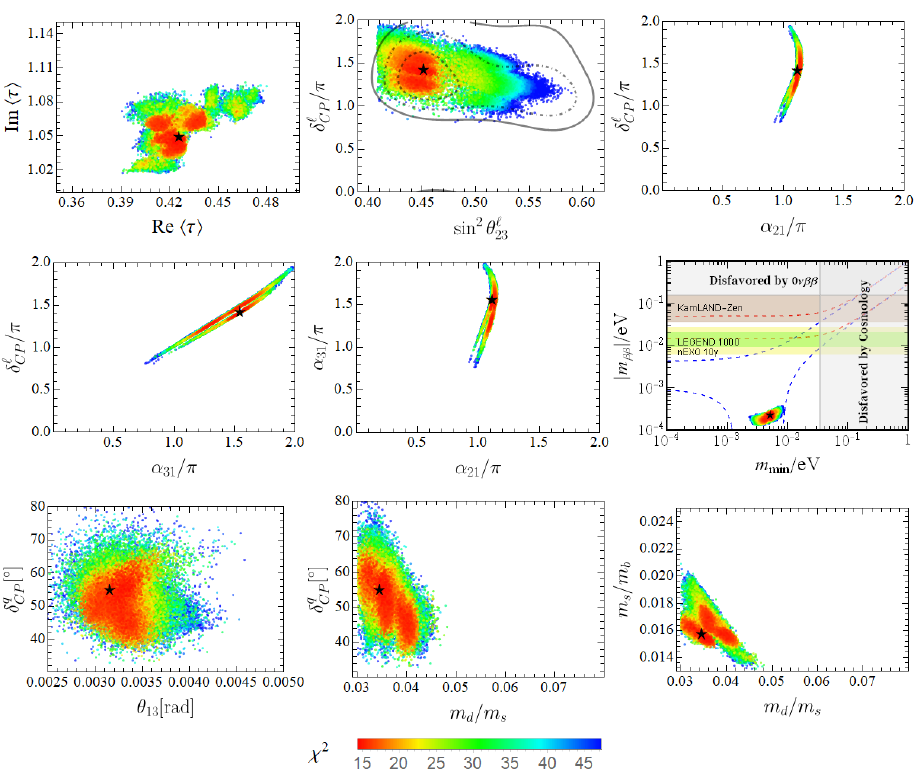}
\caption{\label{fig:model1} The allowed region of $\tau$ and the predictions for the correlations between masses and mixing parameters of quarks and leptons in model 1. The $\chi^2$ is constructed by using the data with SK, and the star ``{\color{black}$\bigstar$}'' refers to the best fitting point. The black dashed line, dash-dotted line and solid line denote the $1\sigma$, $2\sigma$ and $3\sigma$ contours respectively in the $\sin^2\theta^{\ell}_{23}-\delta^{\ell}_{CP}$ plane~\cite{Esteban:2020cvm}. In the panel of the effective Majorana neutrino mass $|m_{\beta\beta}|$ versus the lightest neutrino mass $m_{\text{min}}$, the red (blue) dashed lines indicate the most general allowed regions for normal (inverted) ordering neutrino mass spectrum at $3\sigma$ level~\cite{Esteban:2020cvm}. The light brown, light yellow and light green horizontal bands denote the current experimental bound from KamLAND-Zen~\cite{KamLAND-Zen:2022tow} and the future sensitivity ranges of LEGEND-1000~\cite{LEGEND:2021bnm} and nEXO~\cite{nEXO:2021ujk} respectively. The vertical grey band is excluded by the Planck data $\Sigma_{i}m_{i}<0.120\,\text{eV}$~\cite{Planck:2018vyg}. }
\end{figure}

\begin{figure}[hptb!]
\centering
\includegraphics[width=6.5in]{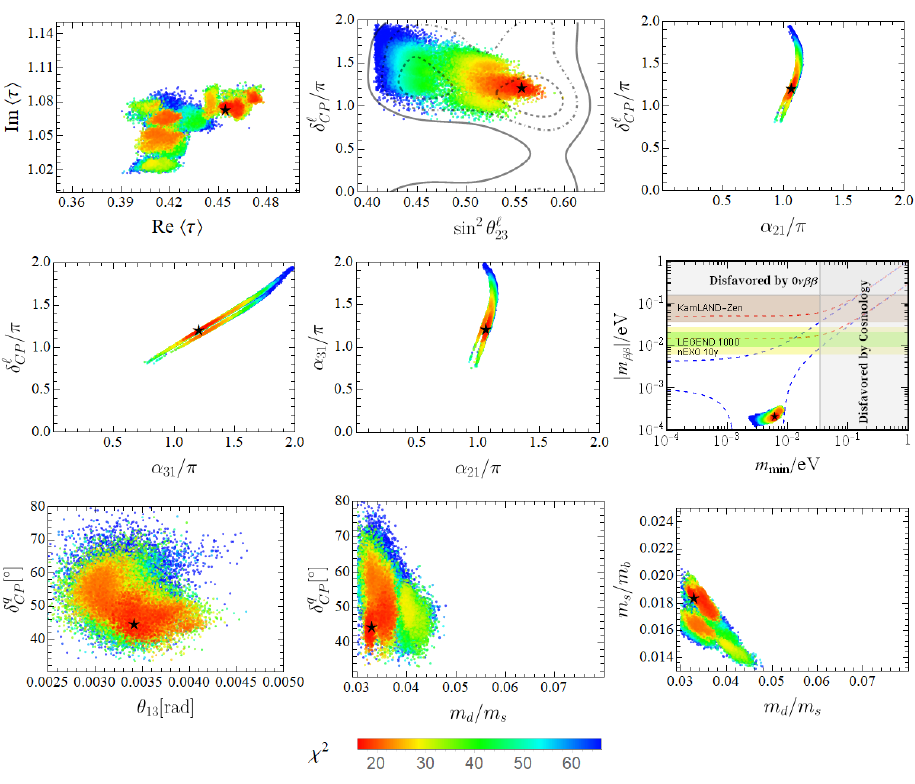}
\caption{\label{fig:model1_withoutSK}
The allowed region of $\tau$ and the predictions for the correlations between masses and mixing parameters of quarks and leptons in model 1. The same convention as figure~\ref{fig:model1} is adopted.  We construct the $\chi^2$ function by using the data without SK. }
\end{figure}

\begin{itemize}
\item{model 1: $(F^c,F_1,F_2,F_3)\sim (\mathbf{3}, \mathbf{1}', \mathbf{1}, \mathbf{1}')$, $(k_{F^c}, k_{F_1}, k_{F_2}, k_{F_3}, k_{\Sigma})=(1, 1, 3, 3, 2)$ }

We can read off the Yukawa superpotential invariant under the modular symmetry as follow,
\begin{eqnarray}
\nonumber\mathcal{W}_{\Phi} &=& \left[ \alpha_1 (F^c F_1)_{\bm{3}} Y^{(2)}_{\bm{3}} + \beta_1 (F^c F_2)_{\bm{3}} Y^{(4)}_{\bm{3}} + \gamma_1 (F^c F_3)_{\bm{3}} Y^{(4)}_{\bm{3}} \right] \Phi\,,  \\
\nonumber \mathcal{W}_{\Sigma} &=& \Big[ \hat{\alpha}_1 (F^c F_1)_{\bm{3}} Y^{(4)}_{\bm{3}} + \hat{\beta}_1 (F^c F_2)_{\bm{3}} Y^{(6)}_{\bm{3}I} + \hat{\beta}_2 (F^c F_2)_{\bm{3}} Y^{(6)}_{\bm{3}II} \\ \nonumber & & ~+ \hat{\gamma}_1 (F^c F_3)_{\bm{3}} Y^{(6)}_{\bm{3}I} + \hat{\gamma}_2 (F^c F_3)_{\bm{3}} Y^{(6)}_{\bm{3}II} \Big] \Sigma \,,  \\
\label{eq:W-model1}\mathcal{W}_{\Delta_R} &=& \alpha_{R1}(F^c F^c)_{\bm{3}} Y^{(2)}_{\bm{3}} \Delta_R \,,
\end{eqnarray}
where the superpotential $\mathcal{W}$ is splitted into three parts $\mathcal{W}_{\Phi}$, $\mathcal{W}_{\Sigma}$ and $\mathcal{W}_{\Delta_R}$ which couple with the Higgs multiplets $\Phi$, $\Sigma$ and $\Delta_R$ respectively. The phases of the couplings $\alpha_1$, $\beta_1$, $\gamma_1$ and $\alpha_{R1}$ can be absorbed by redefinition of matter fields, while all other couplings are complex. Given the decomposition of the tensor product of two $A_4$ triplets in Eq.~\eqref{eq:tensor-product-A4}, we find that the superpotential of Eq.~\eqref{eq:W-model1} leads to the following Yukawa matrices
\begin{eqnarray}
\nonumber {\cal Y}^1&=&
\begin{pmatrix}
\alpha_1 Y^{(2)}_{\bm{3},3} ~&~ \beta_1 Y^{(4)}_{\bm{3},1} ~&~  \gamma_1  Y^{(4)}_{\bm{3},3} \\
\alpha_1 Y^{(2)}_{\bm{3},2} ~&~ \beta_1 Y^{(4)}_{\bm{3},3} ~&~  \gamma_1  Y^{(4)}_{\bm{3},2} \\
\alpha_1 Y^{(2)}_{\bm{3},1} ~&~ \beta_1 Y^{(4)}_{\bm{3},2} ~&~  \gamma_1  Y^{(4)}_{\bm{3},1}
\end{pmatrix}\,, \\
\nonumber {\cal Y}^{15}&=&
\begin{pmatrix}
 \hat{\alpha}_1 Y^{(4)}_{\bm{3},3} ~&~ \hat{\beta}_1 Y^{(6)}_{\bm{3}I,1} + \hat{\beta}_2 Y^{(6)}_{\bm{3}II,1} ~&~ \hat{\gamma}_1 Y^{(6)}_{\bm{3}I,3}+\hat{\gamma}_2 Y^{(6)}_{\bm{3}II,3} \\
 \hat{\alpha}_1 Y^{(4)}_{\bm{3},2} ~&~ \hat{\beta}_1 Y^{(6)}_{\bm{3}I,3} + \hat{\beta}_2 Y^{(6)}_{\bm{3}II,3} ~&~ \hat{\gamma}_1 Y^{(6)}_{\bm{3}I,2}+\hat{\gamma}_2 Y^{(6)}_{\bm{3}II,2}  \\
 \hat{\alpha}_1 Y^{(4)}_{\bm{3},1} ~&~ \hat{\beta}_1 Y^{(6)}_{\bm{3}I,2} + \hat{\beta}_2 Y^{(6)}_{\bm{3}II,2} ~&~ \hat{\gamma}_1 Y^{(6)}_{\bm{3}I,1}+\hat{\gamma}_2 Y^{(6)}_{\bm{3}II,1}
\end{pmatrix}\,, \\
\label{eq:Y-model1}\mathcal{Y}^{10_R}&=&\alpha_{R1}
\begin{pmatrix}
 2 Y^{(2)}_{\bm{3},1} ~&~ -Y^{(2)}_{\bm{3},3} ~&~ -Y^{(2)}_{\bm{3},2} \\
  -Y^{(2)}_{\bm{3},3} ~&~ 2 Y^{(2)}_{\bm{3},2} ~&~ -Y^{(2)}_{\bm{3},1}\\
  -Y^{(2)}_{\bm{3},2} ~&~ -Y^{(2)}_{\bm{3},1} ~&~ 2 Y^{(2)}_{\bm{3},3}
\end{pmatrix}\,.
\end{eqnarray}
This model gives a good fit to the masses and mixing parameters of quarks and leptons, the minimum of the $\chi^2$ function is determined to be $\chi^2_{\text{min}}=14.1395 (15.9917)$ for the neutrino data with (without) SK. The best fit values of the coupling constants and the corresponding values of fermion masses and mixing parameters are summarized in tables~\ref{tab:best-fit-WSK} and \ref{tab:best-fit-WOSK}. Notice that the VEV ratios $\frac{v^u_\Phi}{v_u}$ and $\frac{v^d_\Sigma}{v_u}\frac{v^u_\Phi }{v^d_\Phi}$ have been absorbed into the couplings $\alpha_1$, $\beta_1$, $\gamma_1$ and $\hat{\alpha}_1$,  $\hat{\beta}_1$,  $\hat{\beta}_1$,  $\hat{\gamma}_1$,  $\hat{\gamma}_2$ respectively, as explained in section~\ref{sec:Pati-Salam-GUT-Fmass}. We see that the experimental data of lepton sector can be accommodated very well and the best fit value of $\theta^{\ell}_{23}$ depend on the SK data. It turns out that neutrino masses are normal ordering, certain flavor observables would be outside the experimental $3\sigma$ ranges for inverted ordering. The model can also yield predictions for the unknown Majorana phases $\alpha_{21}$, $\alpha_{31}$ and the absolute values of the light neutrino masses $m_{1, 2, 3}$. The light neutrino mass sum is $\sum_i m_i\approx65.5 (67.6)$ meV which is compatible with the cosmological constraint $\sum_i m_i<120$ meV from Planck~\cite{Planck:2018vyg}. In addition, the effective mass of the neutrinoless double beta ($0\nu\beta\beta$) decay is determined to be $|m_{\beta\beta}|\approx0.2\,\text{meV}$ which is below the sensitivity of the future tonn-scale experiments.

Moreover, we scan the parameter space around the best fit point, and all the observables are required to lie in the experimentally preferred $3\sigma$ regions. The allowed values of the complex modulus $\tau$ and the correlations between the masses and mixing parameters are displayed in figure~\ref{fig:model1} and figure~\ref{fig:model1_withoutSK}, where the best fit values are indicated by black star. In the panel of $\delta^{\ell}_{CP}$ versus $\sin^2\theta^{\ell}_{23}$, the $1\sigma$, $2\sigma$ and $3\sigma$ contours adopted from NuFIT 5.2~\cite{Esteban:2020cvm} are shown in dashed, dash-dotted and solid lines respectively, we see that the atmospheric angle $\theta^{\ell}_{23}$ can be in either the first octant or the second octant in this model. In the panel of the effective Majorana mass $|m_{\beta\beta}|$, the horizontal bands indicate the present experimental bound $|m_{\beta\beta}|<(36-156)\,$meV by  KamLAND-Zen~\cite{KamLAND-Zen:2022tow} and the future sensitivity ranges $|m_{\beta\beta}|<(9.0-21)\,$ meV from LEGEND-1000~\cite{LEGEND:2021bnm} and $|m_{\beta\beta}|<(6.1-27)\,$ meV from nEXO~\cite{nEXO:2021ujk}, and the cosmological bound  $\Sigma_{i}m_{i}<0.120\,\text{eV}$ from Planck~\cite{Planck:2018vyg} is shown by vertical band. The effective mass $|m_{\beta\beta}|$ is quite small so that the signal of $0\nu\beta\beta$ decay is hardly measurable in near future experiments. We would like to remind that the $\chi^2$ function usually has a plenty of local minima in the large parameter space and different predictions could possibly be reached around other local minima.

\item{model 2: $(F^c,F_1,F_2,F_3)\sim (\mathbf{3}, \mathbf{1}', \mathbf{1}, \mathbf{1}')$, $(k_{F^c}, k_{F_1}, k_{F_2}, k_{F_3}, k_{\Sigma})=(0, 2, 6, 8, -2)$ }

This model differs from model 1 in the modular weights of the Higgs and fermion multiplets. The modular invariance fixes the superpotential to be of the following form,
\begin{eqnarray}
\nonumber \mathcal{W}_{\Phi} &=& \Big[ \alpha_1 (F^c F_1)_{\bm{3}} Y^{(2)}_{\bm{3}} + \beta_1 (F^c F_2)_{\bm{3}} Y^{(6)}_{\bm{3}I} + \beta_2 (F^c F_2)_{\bm{3}} Y^{(6)}_{\bm{3}II} \\
\nonumber &&~+ \gamma_1 (F^c F_3)_{\bm{3}} Y^{(8)}_{\bm{3}I} + \gamma_2 (F^c F_3)_{\bm{3}} Y^{(8)}_{\bm{3}II} \Big] \Phi\,, \\
\nonumber \mathcal{W}_{\Sigma} &=& \left[ \hat{\beta}_1 (F^c F_2)_{\bm{3}} Y^{(4)}_{\bm{3}} + \hat{\gamma}_1 (F^c F_3)_{\bm{3}} Y^{(6)}_{\bm{3}I}  + \hat{\gamma}_2 (F^c F_3)_{\bm{3}} Y^{(6)}_{\bm{3}II} \right] \Sigma\,, \\
\mathcal{W}_{\Delta_R} &=& \alpha_{R1}(F^c F^c)_{\bm{1}}  \Delta_R \,.
\end{eqnarray}
The corresponding Yukawa matrices are given by
\begin{eqnarray}
\nonumber {\cal Y}^1&=&
\begin{pmatrix}
\alpha_1 Y^{(2)}_{\bm{3},3}  ~&~  \beta_1 Y^{(6)}_{\bm{3}I,1} + \beta_2 Y^{(6)}_{\bm{3}II,1}  ~&~ \gamma_1  Y^{(8)}_{\bm{3}I,3}+ \gamma_2  Y^{(8)}_{\bm{3}II,3}  \\
\alpha_1 Y^{(2)}_{\bm{3},2}  ~&~  \beta_1 Y^{(6)}_{\bm{3}I,3} + \beta_2 Y^{(6)}_{\bm{3}II,3}  ~&~ \gamma_1   Y^{(8)}_{\bm{3}I,2}+ \gamma_2  Y^{(8)}_{\bm{3}II,2} \\
\alpha_1 Y^{(2)}_{\bm{3},1}  ~&~  \beta_1 Y^{(6)}_{\bm{3}I,2} + \beta_2 Y^{(6)}_{\bm{3}II,2}  ~&~ \gamma_1   Y^{(8)}_{\bm{3}I,1}+ \gamma_2  Y^{(8)}_{\bm{3}II,1}
\end{pmatrix} \,,  \\
\nonumber {\cal Y}^{15}  &=&
\begin{pmatrix}
0 ~&~  \hat{\beta}_1 Y^{(4)}_{\bm{3},1} ~&~  \hat{\gamma}_1 Y^{(6)}_{\bm{3}I,3} + \hat{\gamma}_2 Y^{(6)}_{\bm{3}II,3} \\
0 ~&~  \hat{\beta}_1 Y^{(4)}_{\bm{3},3} ~&~  \hat{\gamma}_1 Y^{(6)}_{\bm{3}I,2} + \hat{\gamma}_2 Y^{(6)}_{\bm{3}II,2} \\
0 ~&~  \hat{\beta}_1 Y^{(4)}_{\bm{3},2} ~&~  \hat{\gamma}_1 Y^{(6)}_{\bm{3}I,1} + \hat{\gamma}_2 Y^{(6)}_{\bm{3}II,1}
\end{pmatrix} \,, \\
\mathcal{Y}^{10_R}&=&\alpha_{R1}
\begin{pmatrix}
  1 ~&~ 0 ~&~ 0 \\
  0 ~&~ 0 ~&~ 1 \\
  0 ~&~ 1 ~&~ 0
\end{pmatrix}\,,
\end{eqnarray}
which take different texture from that of model 1 in Eq.~\eqref{eq:Y-model1}. The first column of ${\cal Y}^{15}$ is vanishing and the masses of the heavy right-handed neutrinos are exactly degenerate in this case. The phases of $\alpha_1$, $\beta_1$, $\gamma_1$ and $\alpha_{R1}$ can be rotated away and all other couplings are complex. The fitting results are listed in table~\ref{tab:best-fit-WSK} and table~\ref{tab:best-fit-WOSK}, the best fits values of fermion masses and the mixing parameters are in excellent agreement with experimental data. The correlations between flavor observables are plotted in figures~\ref{fig:model2} and~\ref{fig:model2_withoutSK}. As can be seen, the effective Majorana mass $|m_{\beta\beta}|$ is predicted to be below the sensitivities of future $0\nu\beta\beta$ decay experiments such as LEGEND 1000 and nEXO 10y.

\end{itemize}

\subsection{$F\sim\mathbf{3}$}

In the phenomenologically viable $\text{PS}\times A_4$ models with minimum number of parameters, one could assign the three generations of left-handed fermions $F_i$ to a $A_4$ triplet $\mathbf{3}$ while $F^c_i$ are singlets of $A_4$. Then the Yukawa couplings of $\Phi$, $\Sigma$ and $\Delta_R$ would be triplet modular forms and singlet modular forms of level 3 respectively. In the following, we present another two viable $\text{PS}\times A_4$ models depending on 19 free parameters under the representation assignment $F\sim\mathbf{3}$ and $F^c_i\sim\mathbf{1}, \mathbf{1}'~\text{or}~ \mathbf{1}''$.

\begin{figure}[hptb!]
\centering
\includegraphics[width=6.5in]{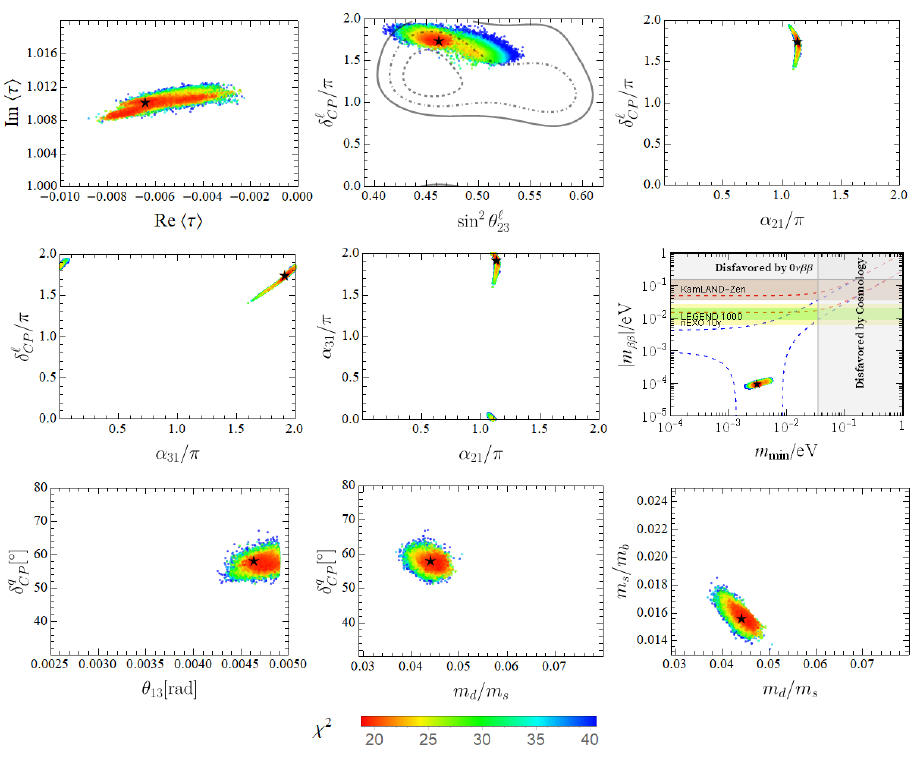}
\caption{\label{fig:model3} The allowed region of $\tau$ and the predictions for the correlations between masses and mixing parameters of quarks and leptons in model 3. The same convention as figure~\ref{fig:model1} is adopted. We construct the $\chi^2$ function by using the data with SK.}
\end{figure}

\begin{figure}[hptb!]
\centering
\includegraphics[width=6.5in]{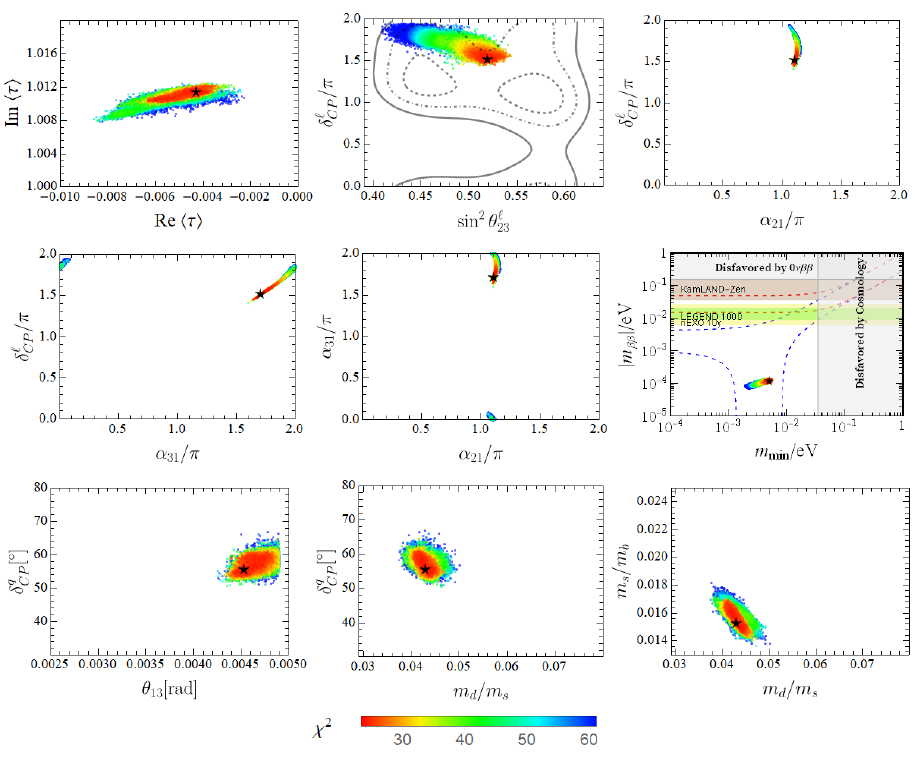}
\caption{\label{fig:model3_withoutSK} The allowed region of $\tau$ and the predictions for the correlations between masses and mixing parameters of quarks and leptons in model 3. The same convention as figure~\ref{fig:model1} is adopted. We construct the $\chi^2$ function by using the data without SK.}
\end{figure}

\begin{figure}[hptb!]
\centering
\includegraphics[width=6.5in]{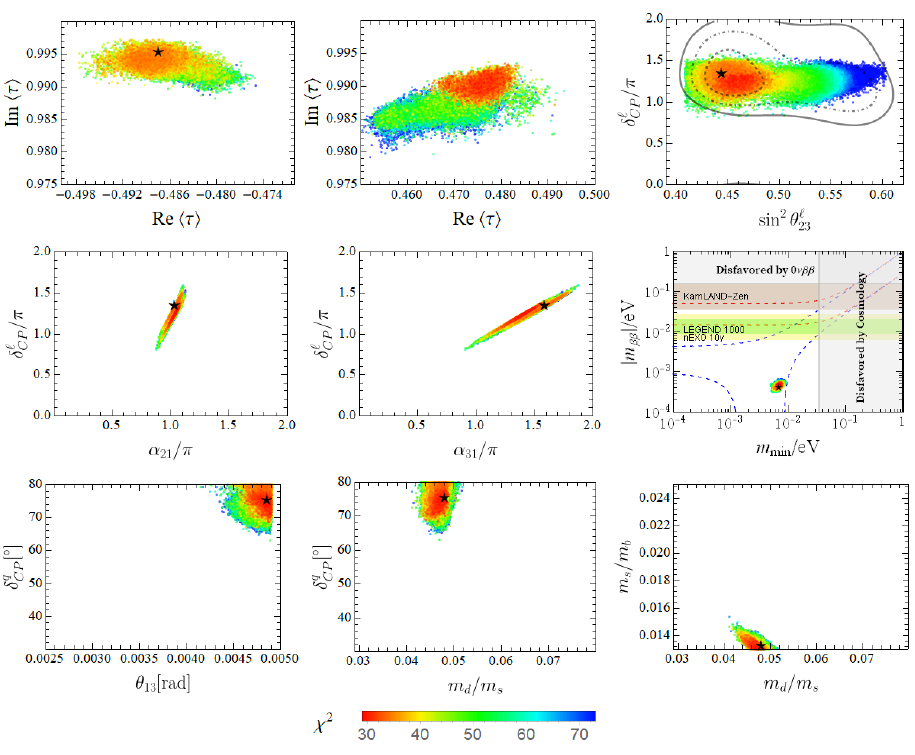}
\caption{\label{fig:model4} The allowed region of $\tau$ and the predictions for the correlations between masses and mixing parameters of quarks and leptons in model 4. The same convention as figure~\ref{fig:model1} is adopted. We construct the $\chi^2$ function by using the data with SK. }
\end{figure}

\begin{figure}[hptb!]
\centering
\includegraphics[width=6.5in]{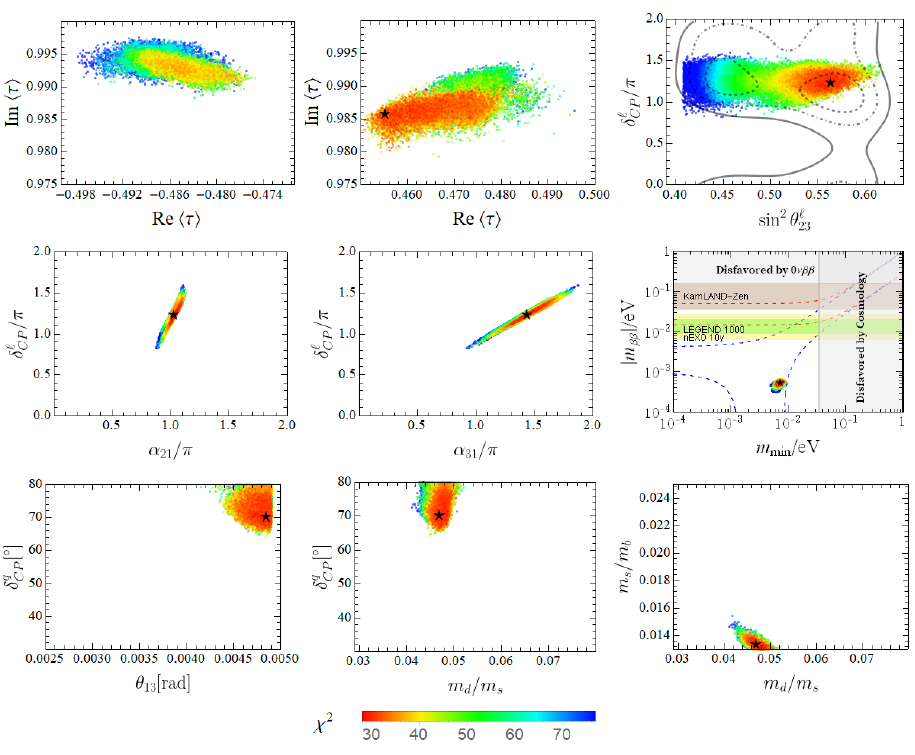}
\caption{\label{fig:model4_withoutSK}The allowed region of $\tau$ and the predictions for the correlations between masses and mixing parameters of quarks and leptons in model 4. The same convention as figure~\ref{fig:model1} is adopted. We construct the $\chi^2$ function by using the data without SK. }
\end{figure}

\begin{itemize}

\item{model 3: $(F^c_1, F^c_2, F^c_3, F) = (\mathbf{1}', \mathbf{1}'', \mathbf{1}', \mathbf{3})$, $(k_{F^c_1}, k_{F^c_2}, k_{F^c_3}, k_{F}, k_{\Sigma})=(0, 0, 2, 2, 2)$ }

The modular invariant superpotential for the masses of quarks and leptons  is given by
\begin{eqnarray}
\nonumber \mathcal{W}_{\Phi} &=& \left[\alpha_1 (F^c_1 F)_{\bm{3}} Y^{(2)}_{\bm{3}} + \beta_1 (F^c_2 F)_{\bm{3}} Y^{(2)}_{\bm{3}} + \gamma_1 (F^c_3 F)_{\bm{3}} Y^{(4)}_{\bm{3}} \right] \Phi \,, \\
\nonumber \mathcal{W}_{\Sigma} &=& \left[ \hat{\alpha}_1 (F^c_1 F)_{\bm{3}} Y^{(4)}_{\bm{3}} + \hat{\beta}_1 (F^c_2 F)_{\bm{3}} Y^{(4)}_{\bm{3}} + \hat{\gamma}_1 (F^c_3 F)_{\bm{3}} Y^{(6)}_{\bm{3}I} + \hat{\gamma}_2 (F^c_3 F)_{\bm{3}} Y^{(6)}_{\bm{3}II} \right] \Sigma \,, \\
\mathcal{W}_{\Delta_R} &=& \left[ \alpha_{R1}(F^c_1 F^c_2)_{\bm{1}} + \alpha_{R2}(F^c_3 F^c_3)_{\bm{1}''} Y^{(4)}_{\bm{1}'} \right] \Delta_R \,,
\end{eqnarray}
from which we can read out the Yukawa matrices as follow,
\begin{eqnarray}
\nonumber {\cal Y}^1 &=&
\begin{pmatrix}
 \alpha_1  Y^{(2)}_{\bm{3},3} ~&~  \alpha_1  Y^{(2)}_{\bm{3},2} ~&~  \alpha_1  Y^{(2)}_{\bm{3},1} \\
 \beta_1  Y^{(2)}_{\bm{3},2} ~&~   \beta_1  Y^{(2)}_{\bm{3},1} ~&~   \beta_1  Y^{(2)}_{\bm{3},3}  \\
 \gamma_1  Y^{(4)}_{\bm{3},3} ~&~   \gamma_1  Y^{(4)}_{\bm{3},2} ~&~   \gamma_1  Y^{(4)}_{\bm{3},1}
\end{pmatrix} \,, \\
\nonumber {\cal Y}^{15} &=&
\begin{pmatrix}
\hat{\alpha}_1  Y^{(4)}_{\bm{3},3} ~&~  \hat{\alpha}_1  Y^{(4)}_{\bm{3},2} ~&~ \hat{\alpha}_1  Y^{(4)}_{\bm{3},1} \\
\hat{\beta}_1  Y^{(4)}_{\bm{3},2} ~&~   \hat{\beta}_1  Y^{(4)}_{\bm{3},1} ~&~  \hat{\beta}_1  Y^{(4)}_{\bm{3},3} \\
\hat{\gamma}_1 Y^{(6)}_{\bm{3}I,3} + \hat{\gamma}_2 Y^{(6)}_{\bm{3}II,3} ~&~ \hat{\gamma}_1 Y^{(6)}_{\bm{3}I,2} + \hat{\gamma}_2 Y^{(6)}_{\bm{3}II,2}   ~&~  \hat{\gamma}_1 Y^{(6)}_{\bm{3}I,1} + \hat{\gamma}_2 Y^{(6)}_{\bm{3}II,1}
\end{pmatrix}\,, \\
\mathcal{Y}^{10_R}&=&
\begin{pmatrix}
0 ~&~ \frac{1}{2}\alpha_{R1} ~&~ 0 \\
\frac{1}{2}\alpha_{R1} ~&~ 0 ~&~ 0 \\
0 ~&~ 0 ~&~ \alpha_{R2} Y^{(4)}_{\bm{1}'}
\end{pmatrix}\,.
\end{eqnarray}
Similar to previous models, the couplings $\alpha_1$, $\beta_1$, $\gamma_1$ and $\alpha_{R1}$ can be set to real. The agreement between the model and the experimental data is optimized for the parameter choice in tables~\ref{tab:best-fit-WSK} and~\ref{tab:best-fit-WOSK}. The corresponding correlations between flavor observables are showed in figure~\ref{fig:model3} and figure~\ref{fig:model3_withoutSK}. Two right-handed neutrinos have a degenerate mass in this model. The data can be accommodated only for normal ordering neutrino masses, and the $0\nu\beta\beta$ effective mass $|m_{\beta\beta}|$ is too small to be measured.

\item{model 4: $(F^c_1, F^c_2, F^c_3, F) = (\mathbf{1}, \mathbf{1}'', \mathbf{1}', \mathbf{3})$, $(k_{F^c_1}, k_{F^c_2}, k_{F^c_3}, k_{F}, k_{\Sigma})=(0, 2, 4, 2, -2)$ }

In this model, the quark and lepton masses are described by the following superpotential invariant under $A_4$ modular symmetry,
\begin{eqnarray}
\nonumber \mathcal{W}_{\Phi} &=& \left[\alpha_1 (F^c_1 F)_{\bm{3}} Y^{(2)}_{\bm{3}} + \beta_1 (F^c_2 F)_{\bm{3}} Y^{(4)}_{\bm{3}} + \gamma_1 (F^c_3 F)_{\bm{3}} Y^{(6)}_{\bm{3}I} + \gamma_2 (F^c_3 F)_{\bm{3}} Y^{(6)}_{\bm{3}II} \right] \Phi \,, \\
\nonumber \mathcal{W}_{\Sigma} &=& \left[ \hat{\beta}_1 (F^c_2 F)_{\bm{3}} Y^{(2)}_{\bm{3}} + \hat{\gamma}_1 (F^c_3 F)_{\bm{3}} Y^{(4)}_{\bm{3}} \right] \Sigma \,, \\
\mathcal{W}_{\Delta_R} &=& \left[ \alpha_{R1}(F^c_1 F^c_1)_{\bm{1}} + \alpha_{R2}(F^c_2 F^c_3)_{\bm{1}} Y^{(6)}_{\bm{1}} + \alpha_{R3}(F^c_3 F^c_3)_{\bm{1}''} Y^{(8)}_{\bm{1}'} \right] \Delta_R \,,
\end{eqnarray}
which give rise to the Yukawa matrices,
\begin{eqnarray}
\nonumber {\cal Y}^1 &=&
\begin{pmatrix}
 \alpha_1  Y^{(2)}_{\bm{3},1} ~&~  \alpha_1  Y^{(2)}_{\bm{3},3} ~&~  \alpha_1  Y^{(2)}_{\bm{3},2} \\
 \beta_1  Y^{(4)}_{\bm{3},2} ~&~   \beta_1  Y^{(4)}_{\bm{3},1} ~&~   \beta_1  Y^{(4)}_{\bm{3},3}  \\
 \gamma_1  Y^{(6)}_{\bm{3},3} + \gamma_2  Y^{(6)}_{\bm{3},3} ~&~   \gamma_1  Y^{(2)}_{\bm{3},2} + \gamma_2  Y^{(6)}_{\bm{3},2} ~&~   \gamma_1  Y^{(2)}_{\bm{3},1} + \gamma_2  Y^{(6)}_{\bm{3},1}
\end{pmatrix}\,,  \\
\nonumber {\cal Y}^{15} &=&
\begin{pmatrix}
0 ~&~  0 ~&~ 0 \\
\hat{\beta}_1  Y^{(2)}_{\bm{3},2} ~&~   \hat{\beta}_1  Y^{(2)}_{\bm{3},1} ~&~  \hat{\beta}_1  Y^{(2)}_{\bm{3},3} \\
\hat{\gamma}_1 Y^{(4)}_{\bm{3}I,3} ~&~ \hat{\gamma}_1 Y^{(4)}_{\bm{3}I,2} ~&~  \hat{\gamma}_1 Y^{(4)}_{\bm{3}I,1}
\end{pmatrix} \,, \\
\mathcal{Y}^{10_R}&=&
\begin{pmatrix}
\alpha_{R1} ~&~ 0 ~&~ 0 \\
0 ~&~ 0 ~&~ \frac{1}{2}\alpha_{R2} Y^{(6)}_{\bm{1}} \\
0 ~&~ \frac{1}{2}\alpha_{R2} Y^{(6)}_{\bm{1}} ~&~ \alpha_{R3} Y^{(8)}_{\bm{1}'}
\end{pmatrix}\,.
\end{eqnarray}
The numerical fitting results are reported in table~\ref{tab:best-fit-WSK} and table~\ref{tab:best-fit-WOSK}, this model can accommodate the experimental data well. The correlation between the fermion masses and mixing parameters are plotted in figure~\ref{fig:model4} and figure~\ref{fig:model4_withoutSK}.

\end{itemize}

\begin{figure}[hptb!]
\centering
\includegraphics[width=6.5in]{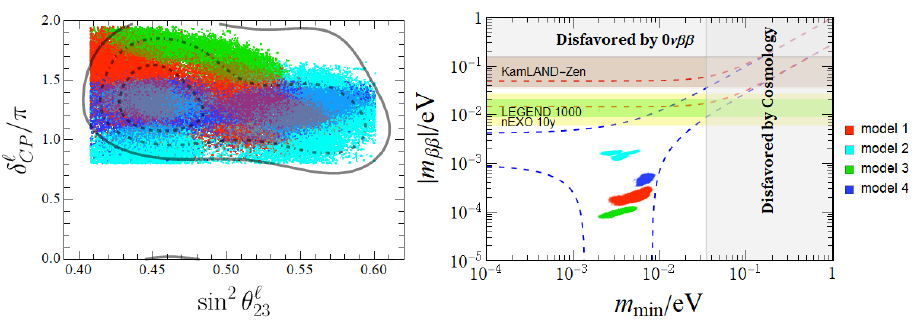}
\caption{\label{fig:collections-lepton-obs-WSK} The predictions for the effective Majorana mass $|m_{\beta\beta}|$ (right panel) and the correlation between $\delta^{\ell}_{CP}$ and $\sin^2\theta^{\ell}_{23}$ (left panel) in the four benchmark PS models based on $A_4$ modular symmetry. We have used the global fit data with SK. }
\end{figure}

\begin{figure}[hptb!]
\centering
\includegraphics[width=6.5in]{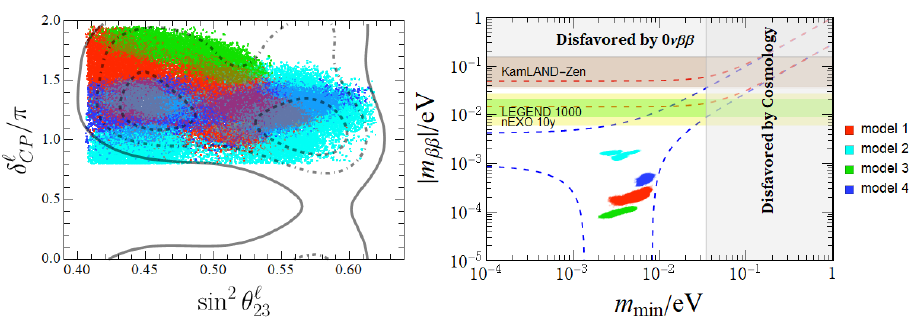}
\caption{\label{fig:collections-lepton-obs-WOSK} The predictions for the effective Majorana mass $|m_{\beta\beta}|$ (right panel) and the correlation between $\delta^{\ell}_{CP}$ and $\sin^2\theta^{\ell}_{23}$ (left panel) in the four benchmark PS models based on $A_4$ modular symmetry. We have used the global fit data without SK.}
\end{figure}

We summarize the predictions for the atmospheric mixing angle $\theta^{\ell}_{23}$, leptonic Dirac CP violation phase $\delta^{\ell}_{CP}$ and the effective mass $|m_{\beta\beta}|$ in figures~\ref{fig:collections-lepton-obs-WSK} and~\ref{fig:collections-lepton-obs-WOSK}. The forthcoming long baseline neutrino experiments DUNE~\cite{DUNE:2015lol} and T2HK~\cite{Hyper-Kamiokande:2018ofw} are expected to be able to precisely measure $\theta^{\ell}_{23}$ and $\delta^{\ell}_{CP}$ so that the four benchmark models would be tested.

\begin{table}[t!]
\centering
\resizebox{1.0\textwidth}{!}{
\begin{tabular}{|c|c|c|c|c|}\hline
 & model 1 & model 2 & model 3 & model 4\\\hline
$\langle \tau \rangle$ & $0.42596 + 1.0490i$ & $0.13624 + 1.7870i$ & $-0.0063859 + 1.0102i$ & $-0.48736 + 0.99525i$  \\
$\beta_1 / \alpha_1$ & $17.150$ & $0.65715$ & $1.0749$ & $0.16838$  \\
$\beta_2 / \alpha_1$ & $-$ & $17.121e^{0.7815\pi i}$ & $-$ & $-$  \\
$\gamma_1 / \alpha_1$ & $1617.9$ & $38.075$ & $282.13$ & $5.2037$  \\
$\gamma_2 / \alpha_1$ & $-$ & $4.7583e^{1.9835\pi i}$ & $-$ & $225.02e^{0.6512\pi i}$  \\
$\hat{\alpha}_1 / \alpha_1$ & $6.9611e^{1.6913\pi i}$ & $-$ & $0.74490e^{1.2525\pi i}$ & $-$  \\
$\hat{\beta}_1 / \alpha_1$ & $84.302e^{0.3081\pi i}$ & $1.2343e^{1.3316\pi i}$ & $102.57e^{1.3675\pi i}$ & $122.14e^{0.0997\pi i}$  \\
$\hat{\beta}_2 / \alpha_1$ & $6.9233e^{0.0737\pi i}$ & $-$ & $-$ & $-$  \\
$\hat{\gamma}_1 / \alpha_1$ & $249.63e^{1.0867\pi i}$ & $24.597e^{0.3783\pi i}$ & $6.9662e^{1.5656\pi i}$ & $11.045e^{0.5141\pi i}$  \\
$\hat{\gamma}_2 / \alpha_1$ & $237.02e^{0.0444\pi i}$ & $3.1842e^{0.0450\pi i}$ & $0.90204e^{0.3969\pi i}$ & $-$  \\
$\alpha_{R2} / \alpha_{R1}$ & $-$ & $-$ & $9.0093e^{0.1331\pi i}$ & $841.24e^{0.4139\pi i}$  \\
$\alpha_{R3} / \alpha_{R1}$ & $-$ & $-$ & $-$ & $8082.3e^{1.3973\pi i}$  \\
$r_2$ & $0.60662e^{0.5275\pi i}$ & $0.53248e^{1.6707\pi i}$ & $1.5138e^{1.6197\pi i}$ & $0.22548e^{0.6019\pi i}$  \\
$\alpha_1 v_u ( {\rm GeV} )$ & $0.0356$ & $1.6025$ & $0.1810$ & $0.1939$  \\
$\alpha_1 r_1 v_d ( {\rm GeV} )$ & $4.0481\times 10^{-4}$ & $0.0166$ & $2.0626\times 10^{-3}$ & $1.9559\times 10^{-3}$  \\
$\frac{\alpha_1^2 v_u^2}{\alpha_{R1} v_R} (\rm meV)$ & $3.7612\times 10^{-3}$ & $6.5100$ & $0.0488$ & $16.3077$  \\
\hline\hline
$\sin^2 \theta^\ell_{12}$ & $0.307$ & $0.301$ & $0.294$ & $0.306$  \\
$\sin^2 \theta^\ell_{13}$ & $0.02211$ & $0.02224$ & $0.02224$ & $0.02163$  \\
$\sin^2 \theta^\ell_{23}$ & $0.452$ & $0.456$ & $0.463$ & $0.444$  \\
$\delta^\ell_{CP} / \pi$ & $1.419$ & $1.295$ & $1.733$ & $1.345$  \\
$\alpha_{21} / \pi$ & $1.119$ & $0.962$ & $1.136$ & $1.039$  \\
$\alpha_{31} / \pi$ & $1.549$ & $0.643$ & $1.917$ & $1.590$  \\\hline
$m_e/m_\mu$ & $0.0047$ & $0.0049$ & $0.0046$ & $0.0045$  \\
$m_\mu/m_\tau$ & $0.060$ & $0.058$ & $0.057$ & $0.058$  \\
$m_1/{\rm meV}$ & $5.158$ & $4.986$ & $3.082$ & $6.989$  \\
$m_2/{\rm meV}$ & $10.035$ & $9.948$ & $9.143$ & $11.088$  \\
$m_3/{\rm meV}$ & $50.326$ & $50.271$ & $50.192$ & $50.577$  \\
$m_{\beta\beta}/{\rm meV}$ & $0.223$ & $1.686$ & $0.094$ & $0.421$  \\\hline
$\theta^q_{12}$ & $0.229$ & $0.229$ & $0.229$ & $0.229$  \\
$\theta^q_{13}$ & $0.0032$ & $0.0035$ & $0.0046$ & $0.0049$  \\
$\theta^q_{23}$ & $0.0392$ & $0.0398$ & $0.0390$ & $0.0386$  \\
$\delta^q_{CP}/\circ$ & $54.76$ & $53.72$ & $58.03$ & $75.29$  \\\hline
$m_u/m_c$ & $0.0027$ & $0.0026$ & $0.0030$ & $0.0028$  \\
$m_c/m_t$ & $0.0028$ & $0.0027$ & $0.0027$ & $0.0029$  \\
$m_d/m_s$ & $0.035$ & $0.038$ & $0.044$ & $0.048$  \\
$m_s/m_b$ & $0.016$ & $0.020$ & $0.016$ & $0.013$  \\\hline
$m_b/m_\tau$ & $0.738$ & $0.755$ & $0.719$ & $0.713$  \\\hline
$\alpha_{R1} v_R/{\rm GeV}$ & $3.378\times 10^{11}$ & $3.945\times 10^{11}$ & $6.715\times 10^{11}$ & $2.305\times 10^{9}$  \\
$M_1/{\rm GeV}$ & $3.120\times 10^{11}$ & $3.945\times 10^{11}$ & $3.358\times 10^{11}$ & $2.305\times 10^{9}$  \\
$M_2/{\rm GeV}$ & $4.965\times 10^{11}$ & $3.945\times 10^{11}$ & $3.358\times 10^{11}$ & $2.179\times 10^{11}$  \\
$M_3/{\rm GeV}$ & $8.085\times 10^{11}$ & $3.945\times 10^{11}$ & $8.630\times 10^{12}$ & $1.571\times 10^{13}$  \\
\hline\hline
$\chi^2_{\ell}$ & $0.7383$ & $0.6306$ & $7.8443$ & $3.1564$  \\
$\chi^2_{q}$ & $13.3255$ & $5.7174$ & $10.8336$ & $28.7200$  \\
$\chi^2_{b\tau}$ & $0.0756$ & $0.6792$ & $0.1303$ & $0.3368$  \\
$\chi^2$ & $14.1395$ & $7.0272$ & $18.8081$ & $32.2132$  \\
\hline\hline
\end{tabular}}
\caption{\label{tab:best-fit-WSK}The best fit values of the free parameters and the corresponding predictions for the fermion masses and mixing parameters in the four benchmark $\text{PS}\times A_4$ modular invariant models. Here we use the global fitting data with SK.}
\end{table}

\begin{table}[t!]
\resizebox{1.0\textwidth}{!}{
\begin{tabular}{|c|c|c|c|c|}\hline
 & model 1 & model 2 & model 3 & model 4\\\hline
$\langle \tau \rangle$ & $0.45442 + 1.0722i$ & $0.33863 + 1.8075i$ & $-0.0042626 + 1.0114i$ & $0.45531 + 0.98576i$  \\
$\beta_1 / \alpha_1$ & $21.226$ & $0.69743$ & $1.1108$ & $0.11561$  \\
$\beta_2 / \alpha_1$ & $-$ & $17.296e^{0.7660\pi i}$ & $-$ & $-$  \\
$\gamma_1 / \alpha_1$ & $1529.2$ & $35.837$ & $294.95$ & $4.9493$  \\
$\gamma_2 / \alpha_1$ & $-$ & $4.5393e^{1.9788\pi i}$ & $-$ & $223.09e^{0.3015\pi i}$  \\
$\hat{\alpha}_1 / \alpha_1$ & $6.3695e^{1.7875\pi i}$ & $-$ & $0.69767e^{1.2649\pi i}$ & $-$  \\
$\hat{\beta}_1 / \alpha_1$ & $76.106e^{0.4174\pi i}$ & $1.1679e^{1.3300\pi i}$ & $103.61e^{1.3840\pi i}$ & $122.03e^{1.4791\pi i}$  \\
$\hat{\beta}_2 / \alpha_1$ & $5.1522e^{0.0621\pi i}$ & $-$ & $-$ & $-$  \\
$\hat{\gamma}_1 / \alpha_1$ & $203.93e^{1.0341\pi i}$ & $23.207e^{0.3844\pi i}$ & $7.2583e^{1.5071\pi i}$ & $10.974e^{0.3484\pi i}$  \\
$\hat{\gamma}_2 / \alpha_1$ & $192.59e^{-0.0048\pi i}$ & $3.0179e^{0.0147\pi i}$ & $0.90427e^{0.3711\pi i}$ & $-$  \\
$\alpha_{R2} / \alpha_{R1}$ & $-$ & $-$ & $9.5103e^{0.0973\pi i}$ & $578.76e^{1.2183\pi i}$  \\
$\alpha_{R3} / \alpha_{R1}$ & $-$ & $-$ & $-$ & $7782.9e^{0.2144\pi i}$  \\
$r_2$ & $0.77953e^{0.5040\pi i}$ & $0.59268e^{1.6718\pi i}$ & $1.5652e^{1.6236\pi i}$ & $0.15763e^{0.7094\pi i}$  \\
$\alpha_1 v_u ( {\rm GeV} )$ & $0.0392$ & $1.6488$ & $0.1736$ & $0.1899$  \\
$\alpha_1 r_1 v_d ( {\rm GeV} )$ & $4.4447\times 10^{-4}$ & $0.0177$ & $1.9857\times 10^{-3}$ & $1.9173\times 10^{-3}$  \\
$\frac{\alpha_1^2 v_u^2}{\alpha_{R1} v_R} (\rm meV)$ & $4.0988\times 10^{-3}$ & $5.2954$ & $0.0573$ & $17.9607$  \\
\hline\hline
$\sin^2 \theta^\ell_{12}$ & $0.307$ & $0.300$ & $0.310$ & $0.308$  \\
$\sin^2 \theta^\ell_{13}$ & $0.02207$ & $0.02229$ & $0.02211$ & $0.02204$  \\
$\sin^2 \theta^\ell_{23}$ & $0.556$ & $0.571$ & $0.520$ & $0.564$  \\
$\delta^\ell_{CP} / \pi$ & $1.202$ & $1.247$ & $1.518$ & $1.235$  \\
$\alpha_{21} / \pi$ & $1.066$ & $0.990$ & $1.113$ & $1.033$  \\
$\alpha_{31} / \pi$ & $1.208$ & $1.688$ & $1.713$ & $1.442$  \\\hline
$m_e/m_\mu$ & $0.0046$ & $0.0048$ & $0.0044$ & $0.0045$  \\
$m_\mu/m_\tau$ & $0.059$ & $0.058$ & $0.057$ & $0.057$  \\
$m_1/{\rm meV}$ & $6.272$ & $3.555$ & $5.025$ & $7.367$  \\
$m_2/{\rm meV}$ & $10.651$ & $9.313$ & $9.968$ & $11.330$  \\
$m_3/{\rm meV}$ & $50.673$ & $50.264$ & $50.093$ & $50.727$  \\
$m_{\beta\beta}/{\rm meV}$ & $0.209$ & $1.333$ & $0.118$ & $0.544$  \\\hline
$\theta^q_{12}$ & $0.229$ & $0.229$ & $0.229$ & $0.229$  \\
$\theta^q_{13}$ & $0.0034$ & $0.0037$ & $0.0045$ & $0.0048$  \\
$\theta^q_{23}$ & $0.0396$ & $0.0397$ & $0.0400$ & $0.0385$  \\
$\delta^q_{CP}/\circ$ & $44.34$ & $59.04$ & $55.61$ & $70.25$  \\\hline
$m_u/m_c$ & $0.0025$ & $0.0029$ & $0.0028$ & $0.0024$  \\
$m_c/m_t$ & $0.0029$ & $0.0027$ & $0.0028$ & $0.0029$  \\
$m_d/m_s$ & $0.033$ & $0.039$ & $0.043$ & $0.047$  \\
$m_s/m_b$ & $0.018$ & $0.021$ & $0.015$ & $0.013$  \\\hline
$m_b/m_\tau$ & $0.751$ & $0.747$ & $0.729$ & $0.717$  \\\hline
$\alpha_{R1} v_R/{\rm GeV}$ & $3.742\times 10^{11}$ & $5.134\times 10^{11}$ & $5.260\times 10^{11}$ & $2.008\times 10^{9}$  \\
$M_1/{\rm GeV}$ & $3.464\times 10^{11}$ & $5.134\times 10^{11}$ & $2.630\times 10^{11}$ & $2.008\times 10^{9}$  \\
$M_2/{\rm GeV}$ & $5.357\times 10^{11}$ & $5.134\times 10^{11}$ & $2.630\times 10^{11}$ & $9.516\times 10^{10}$  \\
$M_3/{\rm GeV}$ & $8.821\times 10^{11}$ & $5.134\times 10^{11}$ & $7.118\times 10^{12}$ & $1.344\times 10^{13}$  \\
\hline\hline
$\chi^2_{\ell}$ & $2.0904$ & $1.0418$ & $14.0982$ & $3.9491$  \\
$\chi^2_{q}$ & $13.4108$ & $4.2982$ & $10.8920$ & $24.5670$  \\
$\chi^2_{b\tau}$ & $0.4905$ & $0.3091$ & $0.0004$ & $0.1919$  \\
$\chi^2$ & $15.9917$ & $5.6491$ & $24.9906$ & $28.7081$  \\
\hline\hline
\end{tabular}}
\caption{\label{tab:best-fit-WOSK}The best fit values of the free parameters and the corresponding predictions for the fermion masses and mixing parameters in the four benchmark $\text{PS}\times A_4$ modular invariant models. Here we use the global fitting data without SK. }
\end{table}

\section{Conclusion and summary}
\label{sec:conclusion}

In this paper we have studied the Pati-Salam (PS) model based on the gauge symmetry $SU(4)_C\times SU(2)_L\times SU(2)_R$, in the framework of $A_4$ modular flavor symmetry.
The flavor symmetry relates the three family of quark/lepton fields, and the gauge symmetry acts on the different fields of one family within a gauge multiplet. In general, both flavor symmetry and GUTs are powerful theoretical tools to address the flavor puzzle, and their combination can lead to more structure in the mass matrices of quark and lepton than considering them alone. Modular flavor symmetry can alleviate the complex vacuum alignment problem of traditional flavor symmetry and the Yukawa couplings are modular forms which depend on the single complex modulus $\tau$, thus modular flavor models are simplified significantly.

The choice of the Pati-Salam gauge symmetry $SU(4)_C\times SU(2)_L\times SU(2)_R$ is preferred to $SO(10)$ in many constructions inspired by string theory at high energy scale, as the gauge symmetry is broken by Higgs multiplets in small dimensional representations which readily arise in string constructions. Since both modular symmetry and PS are motivated from top-down string constructions, this motivates the present study. In comparison with the PS models based on traditional flavor symmetries, both flavons and driving fields are absent so that the modular invariant PS models are much more simpler, with $A_4$ being the minimal modular group which admits triplet representations.

In the PS theories, all fifteen standard model fermions in each family plus a right-handed neutrino are unified into two PS multiplets $F_i\sim (\bm{4}, \bm{2}, \bm{1})$ and $F^c_i\sim (\overline{\bm{4}}, \bm{1}, \overline{\bm{2}})$. In the present paper, we consider the minimal scenario that the light neutrino mass is generated through the type-I seesaw mechanism. Two PS Higgs multiplets $\Phi \sim (\bm{1},\bm{2},\bm{2})$, $\Sigma\sim (\bm{15},\bm{2},\bm{2})$ are introduced to trigger the electroweak symmetry breaking, and  $\Sigma$ can lead to the Georgi-Jarlskog factor $-3$ so that the difference in the mass spectrum of down type quarks and charged leptons can be accommodated. After including the $A_4$ modular symmetry, one can assign the three generations of PS matter multiplets $F_i$ and $F^c_i$ to be triplet of singlets under the action of $A_4$. For instance, both $F_i$ and $F^c_i$ can transform as irreducible triplet under $A_4$, $F_i$(or $F^c_i$) form an $A_4$ triplet while $F^c_i$(or $F_i$) are $A_4$ singlets. The fermion mass matrices for each possible assignment. However, we have dropped the assignment that both $F_i$ and $F^c_i$ are $A_4$ singlets, since the flavor symmetry would be essentially the $Z_3$ subgroup generated by modular generator $T$ and the different entries of the Yukawa couplings would be loosely correlated for this kind of assignment.

We have performed a comprehensive numerical analysis of all such $\text{PS}\times A_4$ modular models. We find that the phenomenologically viable models involves at least 19 real parameters including the real and imaginary parts of $\tau$. Furthermore, we present four benchmark models in which either $F_i$ or $F^c_i$ transforms as triplet $\mathbf{3}$ under $A_4$ modular symmetry. All the four minimal models depend on 19 real parameters and they provide an excellent description of quark and lepton masses, mixing and CP violation. The best fitting points and the corresponding values of flavor observables are listed in tables~\ref{tab:best-fit-WSK} and~\ref{tab:best-fit-WOSK}. We perform a numerical scan for each model, and plot the allowed region of $\tau$ and the correlations between the fermion masses and mixing parameters. The neutrino mass spectrum is determined to be normal ordered in all these four models, since the best fit values of certain flavor observables lie outside the $3\sigma$ regions for inverted ordering.
The predictions for the effective Majorana mass $|m_{\beta\beta}|$ and the correlation between $\delta^{\ell}_{CP}$ and $\sin^2\theta^{\ell}_{23}$
of all four benchmark models are summarised in figures~\ref{fig:collections-lepton-obs-WSK} and~\ref{fig:collections-lepton-obs-WOSK}.

The predictions of the normal neutrino mass ordering, the atmospheric mixing angle $\theta^{\ell}_{23}$ and leptonic Dirac CP violation phase $\delta^{\ell}_{CP}$ are expected to be either confirmed or excluded by the forthcoming neutrino facilities such as the DUNE and T2HK. The effective mass $|m_{\beta\beta}|$ of the neutrinoless double beta decay is
expected to be in the range of 0.1-2 meV which is below the sensitivities of the future tonn-scale experiments.

%%%%%%%%%%%%%%%%%%%%%%%%%%%%%%%%%%%%%%%%%%%%%
\section*{Acknowledgements}

GJD is supported by the National Natural Science Foundation of China under Grant Nos.~12375104, 11975224. JNL is supported by the Grants No. NSFC-12147110 and the China Post-doctoral Science Foundation under Grant No. 2021M70. SFK would like to thank CERN for its hospitality. SFK acknowledges the STFC Consolidated Grant ST/L000296/1 and the European Union's Horizon 2020 Research and Innovation programme under Marie Sk\l{}odowska-Curie grant agreement HIDDeN European ITN project (H2020-MSCA-ITN-2019//860881-HIDDeN).

\clearpage

%%%%%%%%%%%%%%%%%%%%%%%%%%%%%%%%%%%%%%%%%%%%%

\section*{Appendix}

\setcounter{equation}{0}
\renewcommand{\theequation}{\thesection.\arabic{equation}}

\begin{appendix}

\section{\label{app:A4-MDF} Finite modular group $\Gamma_3\cong A_4$ and higher weight modular multiplets of level 3 }

The inhomogeneous finite modular group $\Gamma_3$ is isomorphic to the permutation group $A_4$, the generators of $A_4$ can be chosen as $S$ and $T$ obeying the relations $S^2=(ST)^3=T^3=1$. The twelve elements of $A_4$ can be arranged into four conjugacy classes,
\begin{align}
\nonumber 1C_1&=\left\{S^2\right\}\,,\\
\nonumber 3C_2&=\left\{S, TST^2, T^2ST\right\}\,,\\
\nonumber 4C_3&=\left\{T, ST, TS, STS\right\}\,,\\
4C'_3&=\left\{T^2, ST^2, T^2S, ST^2S\right\}\,.
\end{align}
The number of irreducible representations is equal to the number of conjugacy classes for a finite group. Hence the $A_4$ group has four irreducible representations: three singlets $\bm{1}$, $\bm{1}'$, $\bm{1}''$ and one triplet $\bm{3}$. The generator $T$ is represented by $1$, $\omega$ and $\omega^2$ respectively in the singlet representations $\bm{1}$, $\bm{1}'$ and $\bm{1}''$ while the generator $S$ is 1, where $\omega=e^{2\pi i/3}$. We work in the $T-$diagonal basis in which the representation matrices of the generators are
\begin{align}
\bm{3}&:~ S=\frac{1}{3}\left(\begin{array}{ccc}
 -1 ~&~ 2  ~&~ 2  \\
  2  ~&~ -1  ~&~ 2 \\
  2 ~&~ 2 ~&~ -1
\end{array}
\right)\,, ~~~~
T=\left(\begin{array}{ccc}
 1 ~&~ 0 ~&~ 0 \\
 0 ~&~ \omega ~&~ 0 \\
 0 ~&~ 0 ~&~ \omega^{2}
\end{array}
\right) \,.
\end{align}
The products of singlet representations are
\begin{equation}
\mathbf{1}'\otimes\mathbf{1}'=\mathbf{1}''\,,~~~\mathbf{1}''\otimes\mathbf{1}''=\mathbf{1}'\,,~~~\mathbf{1}'\otimes\mathbf{1}''=\mathbf{1}\,.
\end{equation}
The decomposition of the tensor product of two $A_4$ triplets is $\bm{3}\otimes\bm{3}=\bm{1}\oplus \bm{1}'\oplus\bm{1}''\oplus\bm{3}_S\oplus\bm{3}_A$. Given two triplets $\alpha=(\alpha_1, \alpha_2, \alpha_3)$ and $\beta=(\beta_1, \beta_2, \beta_3)$, one has the following contractions rules
\begin{align}
\nonumber\bm{1}&\equiv\left(\alpha\beta\right)_{\bm{1}}=\alpha_1\beta_1+\alpha_2\beta_3+\alpha_3\beta_2\,,\\
\nonumber\bm{1}'&\equiv\left(\alpha\beta\right)_{\bm{1}'}=\alpha_3\beta_3+\alpha_1\beta_2+\alpha_2\beta_1\,,\\
\nonumber\bm{1}''&\equiv\left(\alpha\beta\right)_{\bm{1}''}=\alpha_2\beta_2+\alpha_1\beta_3+\alpha_3\beta_1\,, \\
\nonumber\bm{3}_S&\equiv\left(\alpha\beta\right)_{\bm{3}_S}=\left( \begin{array}{c}
2\alpha_1\beta_1-\alpha_2\beta_3-\alpha_3\beta_2 \\
2\alpha_3\beta_3-\alpha_1\beta_2-\alpha_2\beta_1\\
2\alpha_2\beta_2-\alpha_1\beta_3-\alpha_3\beta_1
\end{array}\right)\,,\\
\label{eq:tensor-product-A4} \bm{3}_A&\equiv\left(\alpha\beta\right)_{\bm{3}_A}=\left( \begin{array}{c}
\alpha_2\beta_3-\alpha_3\beta_2\\
\alpha_1\beta_2-\alpha_2\beta_1\\
\alpha_3\beta_1-\alpha_1\beta_3
\end{array}\right)\,.
\end{align}
where $\bm{3}_S$ and $\bm{3}_A$ denote the symmetric and antisymmetric combinations respectively.

\subsection{Modular multiplets of higher weight at level 3 }

Using the tensor product decomposition in Eq.~\eqref{eq:tensor-product-A4}, one can express the higher weight modular weights as polynomial of $Y_1(\tau)$, $Y_2(\tau)$ and $Y_3(\tau)$. The explicit expressions of higher weight modular forms of level 3 have already been given in the literature, for instance in Ref.~\cite{Ding:2021eva}. However, we would like to give them here for self-containess, and the linearly independent modular multiplets up to weight 10 are summarized in table~\ref{tab:modular_forms_A4}.

\begin{table}[t!]
\centering
\begin{tabular}{|c|c|}\hline
Modular weight $k$ & Modular form $Y_{\mathbf{r}}^{(k)}$ \\\hline
  $k=2$ &  $Y^{(2)}_{\bm{3}}=
\begin{pmatrix}
Y_1\\
Y_2\\
Y_3
\end{pmatrix}\,.$ \\ \hline

  \multirow{5}{*}{$k=4$} & $Y^{(4)}_{\bm{1}}=Y_1^2+2 Y_2 Y_3\,,$ \\
  & $Y^{(4)}_{\bm{1}'}=Y_3^2+2 Y_1 Y_2\,,$ \\
  & $Y^{(4)}_{\bm{3}}=
\begin{pmatrix}
Y_1^2-Y_2 Y_3\\
Y_3^2-Y_1 Y_2\\
Y_2^2-Y_1 Y_3
\end{pmatrix}\,.$ \\ \hline
  \multirow{7}{*}{$k=6$} & $Y^{(6)}_{\bm{1}}=Y_1^3+Y_2^3+Y_3^3-3 Y_1 Y_2 Y_3\,,$ \\
  & $Y^{(6)}_{\bm{3}I}=(Y_1^2+2Y_2Y_3)\begin{pmatrix}
Y_1\\
Y_2\\
Y_3
\end{pmatrix}\,,$\\
  & $Y^{(6)}_{\bm{3}II}=
(Y_3^2+2 Y_1Y_2)\begin{pmatrix}
Y_3\\
Y_1\\
Y_2
\end{pmatrix}\,.$ \\ \hline
\multirow{9}{*}{$k=8$} & $Y^{(8)}_{\bm{1}}=(Y_1^2+2 Y_2 Y_3)^2\,,$\\
& $Y^{(8)}_{\bm{1'}}=(Y_1^2+2 Y_2 Y_3)(Y_3^2+2 Y_1 Y_2)\,,$ \\
& $Y^{(8)}_{\bm{1''}}=(Y_3^2+2 Y_1 Y_2)^2\,,$\\
& $Y^{(8)}_{\bm{3}I}=(Y_1^3+Y_2^3+Y_3^3-3 Y_1 Y_2 Y_3)\begin{pmatrix}
Y_1 \\
Y_2\\
Y_3
\end{pmatrix}\,,$ \\
& $Y^{(8)}_{\bm{3}II}=(Y_3^2+2 Y_1Y_2)\begin{pmatrix}
Y^2_2-Y_1Y_3\\
Y^2_1-Y_2Y_3\\
Y^2_3-Y_1Y_2
\end{pmatrix}\,.$\\ \hline
\multirow{11}{*}{$k=10$} & $Y^{(10)}_{\bm{1}}=(Y_1^2+2 Y_2 Y_3)(Y_1^3+Y_2^3+Y_3^3-3 Y_1 Y_2 Y_3)\,,$\\
& $Y^{(10)}_{\bm{1'}}=(Y_3^2+2 Y_1 Y_2)(Y_1^3+Y_2^3+Y_3^3-3 Y_1 Y_2 Y_3)\,,$ \\
& $Y^{(10)}_{\bm{3}I}=(Y_1^2+2 Y_2 Y_3)^2\begin{pmatrix}
Y_1 \\
Y_2\\
Y_3
\end{pmatrix}\,,$\\
& $Y^{(10)}_{\bm{3}II}=(Y_1^2+2 Y_2 Y_3)(Y_3^2+2 Y_1 Y_2)\begin{pmatrix}
Y_3 \\
Y_1\\
Y_2
\end{pmatrix}\,,$\\
& $Y^{(10)}_{\bm{3}III}=(Y_3^2+2 Y_1 Y_2)^2\begin{pmatrix}
Y_2 \\
Y_3\\
Y_1
\end{pmatrix}\,.$\\
  \hline\hline
\end{tabular}
\caption{\label{tab:modular_forms_A4}The summary of modular forms of level 3 with weight up to $10$.}
\end{table}

\end{appendix}

\clearpage

\begin{figure}[hptb!]
\centering
\includegraphics[width=6.5in]{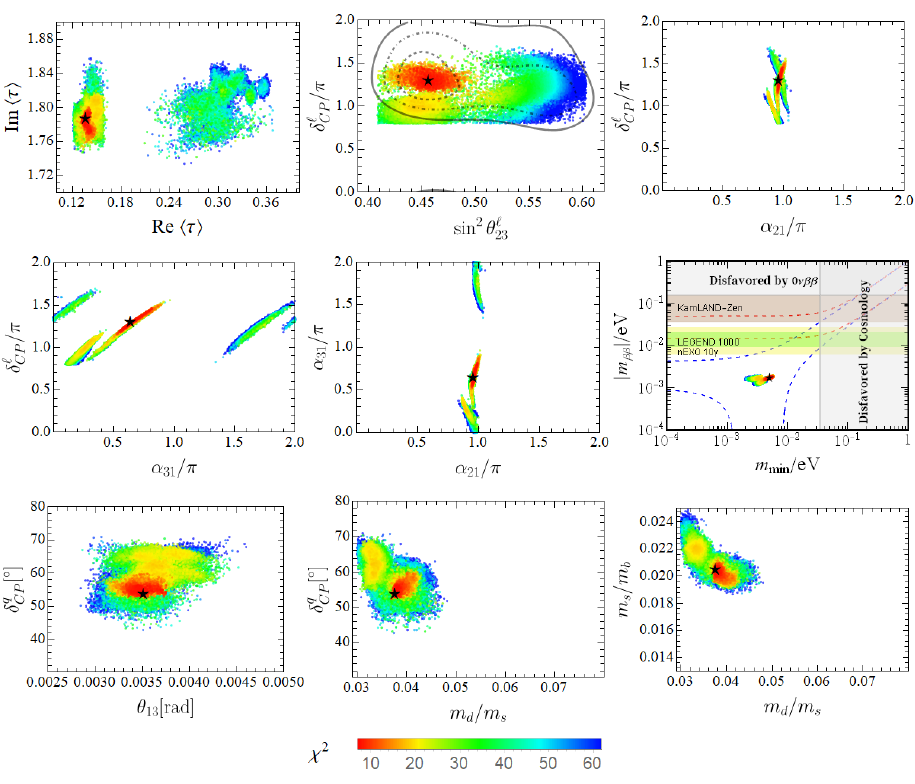}
\caption{\label{fig:model2}
The allowed region of $\tau$ and the predictions for the correlations between masses and mixing parameters of quarks and leptons in model 2. The same convention as figure~\ref{fig:model1} is adopted. We construct the $\chi^2$ function by using the data with SK. }
\end{figure}

\begin{figure}[hptb!]
\centering
\includegraphics[width=6.5in]{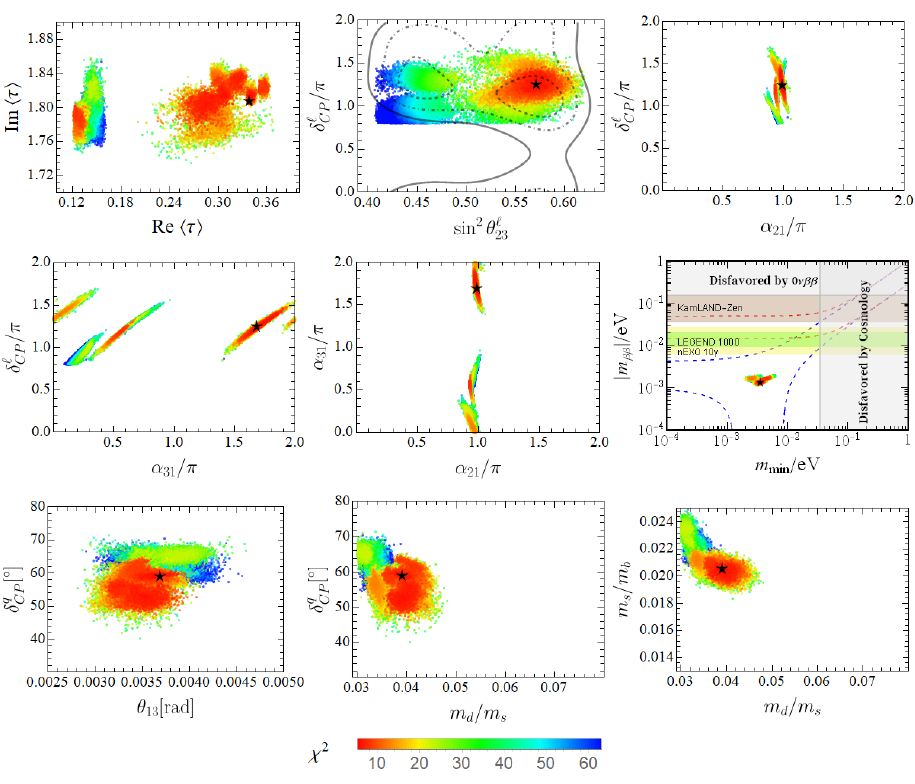}
\caption{\label{fig:model2_withoutSK} The allowed region of $\tau$ and the predictions for the correlations between masses and mixing parameters of quarks and leptons in model 2. The same convention as figure~\ref{fig:model1} is adopted. We construct the $\chi^2$ function by using the data without SK.}
\end{figure}

\clearpage

%\bibliographystyle{utphys}
%\bibliography{references}
%\end{document}

\providecommand{\href}[2]{#2}\begingroup\raggedright\endgroup

\end{document}